\documentclass{aa}  

\usepackage{float}
\usepackage{graphicx}
\usepackage{txfonts}
\usepackage{color}

\usepackage{soul}   %% use \st{}

\begin{document}

   \title{The M101 group complex: new dwarf galaxy candidates and spatial structure}

   \author{Oliver M\"uller
   \inst{1}
          \and
          Roberto Scalera
          \inst{1}
          \and
          Bruno Binggeli
          \inst{1}
          \and
          Helmut Jerjen
          \inst{2}
          }

\institute{Departement Physik, Universit\"at Basel, Klingelbergstr. 82, CH-4056 Basel, Switzerland\\
         \email{oliver89.mueller@unibas.ch; roberto.scalera@stud.unibas.ch; bruno.binggeli@unibas.ch}
         \and
Research School of Astronomy and Astrophysics, Australian National University, Canberra, ACT 2611, Australia\\
         \email{helmut.jerjen@anu.edu.au}
             }

   \date{Received; accepted:}

% \abstract{}{}{}{}{} 
% 5 {} token are mandatory
 
 \abstract
 {The fine details of the large-scale structure in the local universe provide important empirical benchmarks for testing cosmological models of structure formation. Dwarf galaxies are key object for such studies.}
   {Enlarge the sample of known dwarf galaxies in the local universe. We performed a search for faint, unresolved low-surface brightness dwarf galaxies in the M\,101 group complex, including the region around the major spiral galaxies M\,101, M\,51, and M\,63 lying at a distance 7.0, 8.6, and 9.0 Mpc, respectively. The new dwarf galaxy sample can be used in a first step to test for significant substructure in the 2D-distribution and in a second step to study the spatial distribution of the galaxy complex.}
   {Using filtering algorithms we surveyed 330 square degrees of imaging data obtained from the Sloan Digital Sky Survey. The images were visually inspected. The spatial distribution of known galaxies and candidates was analyzed transforming the system into a M\,101 eigenframe, using the geometrical alignment of the group.}
   {We discovered 15 new dwarf galaxies and carried out surface photometry in the $g$ and $r$ bands. The similarity of the photometric properties of these dwarfs to those of Local Group dwarfs suggest membership to the M\,101 group complex. The sky distribution of the candidates follows the thin planar structure outlined by the known members of the three subgroups. The $\sim$ 3\,Mpc long filamentary structure has a $rms$ thickness of 67\,kpc. The planar structure of the embedded M\,101 subgroup is even thinner, with $rms=46$\,kpc. The formation of this structure might be due to the expansion of the Local Void to which it borders.
   Other implications are discussed as well. }
 {We show the viability of SDSS data to extend the sample of dwarfs in the local universe and test cosmological models on small scales.}


   \maketitle
%
%________________________________________________________________
\section{Introduction}

Searching the night sky for new stellar systems of ever lower luminosity and surface brightness, with the aim to enlarge the census of known galaxies in the Local Volume \citep[LV, $D\leq$ 10 Mpc,][]{2013AJ....145..101K}, is a permanent and important task of extragalactic astronomy. Aside from an assessment of the faint-end slope of the galaxy luminosity function \citep[e.g.][]{2002MNRAS.335..712T}, which is a key observation for models of galaxy formation and evolution, it is above all the study of the ``fine structure of large-scale structure'' \citep{1989ASSL..151...47B}, i.e.\,the 3D distribution of low-mass galaxies on large scales, that is fed and fostered by the detection of new nearby dwarf galaxies. Low-mass galaxies are expected to trace the distribution of non-baryonic dark matter (DM) on scales from kpc to Mpc, thus serving as a major testbed for models of structure formation. This is highlighted by the recent discovery of surprisingly thin planes of dwarf satellites in the Local Group (LG), around the Milky Way and the Andromeda galaxy \citep{2012MNRAS.423.1109P,2013MNRAS.435.1928P,2013Natur.493...62I}. The significance of these structures in the context of $\Lambda$CDM or alternative cosmological models is hotly debated  (\citealp{2012PASA...29..395K}, \citealp{2015ApJ...815...19P} $versus$ \citealp{2014MNRAS.443.1274L}, \citealp{2015MNRAS.452.3838C}). But what is not debated is the urgent need to test other nearby groups of galaxies for the existence of similar features, as the ubiquity of the phenomenon would be a challenge to the standard $\Lambda$CDM scenario of structure formation. 

The well-known groups of galaxies in the LV (e.g.~the Local Group, M81\,Group, Cen A/M83 Group, IC 342/Maffei Group, Sculptor filament, and Canes Venatici cloud) have been, and are being, searched for new dwarfs to various surface brightness depths. Three surveys were recently conducted in the southern hemisphere in the directions of the loose Sculptor filament and the rich Centaurus group. There is (1) the very deep but spatially limited {(15 deg$^2$)} PISCeS survey \citep{2014ApJ...793L...7S, 2014ApJ...795L..35C,2016ApJ...823...19C}, (2) the Dark Energy Survey Camera (DECam) based SCABS survey {(21 deg$^2$)} in five photometric bands $ugriz$ \citep{2016arXiv160807285T,2016arXiv160807288T}, and (3) our own 550 deg$^2$ DECam large-field survey \citep{2015A&A...583A..79M,2017A&A...597A...7M}, resulting in the discovery of dozens of new dwarf galaxies. One of our candidates, dw1335-29, has already been confirmed using the TRGB method \citep{2017MNRAS.465.5026C}, more is to follow (M\"uller et al. in preparation). \cite{2015ApJ...802L..25T} reported two almost parallel satellite planes in the Centaurus\,A group. However, with the detection of multiple new dwarf galaxies around Cen\,A this bimodal structure is called into question now \citep{2016A&A...595A.119M}. 
In the northern hemisphere, dedicated deep searches, resulting in the detection of numerous new dwarfs down to a completeness limit of $M_R \approx -10$, were carried out in the rich M\,81 group by \cite{2009AJ....137.3009C, 2013AJ....146..126C}. The authors noted that the satellites lie in a flattened (though not planar) distribution. In the M\,101 group, which is the focus of the present study, the Dragonfly telescope \citep{2014ApJ...787L..37M} and an amateur collective \citep{2016A&A...588A..89J} detected eight new dwarf candidates. Both surveys were confined to the immediate vicinity of M\,101 (9 deg$^2$), leaving out a large portion of the M\,101 group complex that includes M\,51 and M\,63 (see below). A recent HST follow-up of the seven Dragonfly dwarf candidates revealed that four candidates (M101-DF4-7) are in fact ultra diffuse galaxies most likely associated with a background group containing the ellipticals NGC5485 and NGC5473 at a distance of $\sim$27\,Mpc \citep{2016ApJ...833..168M}. 

Surprisingly, two out of the three new faint Dragonfly dwarf members of the M101 group {\citep{2017arXiv170204727D}} are also visible on images of the shallower Sloan Digital Sky Survey (SDSS, www.sdss.org). Equally surprising, with few exceptions \citep{2004AJ....127..704K}, the SDSS has not been employed for systematic searches for unresolved, low-surface brightness dwarf galaxies over a large sky area. We therefore decided to hunt for new dwarfs in a large SDSS region of 330 square degrees covering not only the M\,101 group, but the smaller, neighbouring groups around M\,51 and M\,63 as well, which seem to be connected to the former in a filamentary structure (see Fig.\,3 in \citealp{2013AJ....146...69C}, also Fig.\,\ref{fieldImage} below) -- a structure which we tentatively call here the ``M\,101 group complex''.     

The M\,101 group is more distant at $ 6.95$\,Mpc \citep[][]{2015MNRAS.449.1171N, 2013AJ....145..101K} when compared to the rich M\,81 and Centaurus A groups (at 4-5 Mpc), and is completely dominated by the bulgeless spiral galaxy M\,101. The group is known for its lack of low mass galaxies and is possibly the poorest group in the LV \citep{1999A&AS..137..337B}.{ Eleven out of 14 confirmed members of the M\,101 group complex are late-type spirals and dwarf irregular (dIrr) galaxies -- KK\,191, NGC\,5023, DDO\,182, Holm\,IV, NGC\,5474, NGC\,5477, KKH\,87, DDO\,194 \citep{2013AJ....145..101K}, NGC\,5195 \citep{2001ApJ...546..681T},  DF1 \citep{2017arXiv170204727D}, NGC\,5585 \citep{1994BSAO...38....5K}. Only one is an early-type dwarf elliptical (dE) \citep[UGC\,08882;][]{2005A&A...437..823R} and two are dwarf spheoridal (dSph) galaxies \citep[DF2, DF3;][]{2017arXiv170204727D}.} This stands in direct contrast to rich groups and clusters where {early-type dwarf} galaxies are the most abundant type of galaxies \citep{1987AJ.....94..251B}. The neighbouring and environmentally related spiral galaxies M51 \citep[8.6\,Mpc;][]{2016ApJ...826...21M} and M\,63 \citep[9\,Mpc;][]{2009AJ....138..332J} with their entourage are slightly farther away. It has been debated whether M\,51 and M\,63 plus satellites should be counted as members of the M\,101 group. \citet{2015AstL...41..239T} argue against this view. In the present work, based on our analysis of the galaxy distribution in the region, we will use the term M\,101 group complex for all three galaxies and their satellites and the term subgroup for an individual host and its satellite population (M\,101 subgroup, M\,51 subgroup, and M\,63 subgroup). 

In the first part of the paper (Sections 2-4) we present our search for new low-surface brightness dwarfs in the region of the M\,101 group complex with publicly available SDSS data. We report the discovery of 15 dwarf candidates and perform standard $r$ and $g$ surface photometry for these. As shown in Sect.\,6.1, the photometric parameters of most candidates do suggest galaxy membership in the complex. In the second part of the paper (Sections 5, 6.2) we study the structure of the M\,101 group complex by introducing a suitable reference frame (Sect.\,5). By a cosmic coincidence it happens that the best-fitting plane through the M\,101 subgroup members with known distances is almost seen edge-on with respect to our line of sight (similar to the Centaurus group, \citealp{2016A&A...595A.119M}). This allows a first assessment of where the new candidates lie in the complex without distance information. The filamentary or planar structure of the M\,101 group complex is critically discussed in Sect.\,6.2, followed by a general conclusion in Sect.\,7. 

\section{SDSS Data}
\begin{figure*}[ht]
\centering
\hspace*{-0.7cm}
\includegraphics[width=14cm]{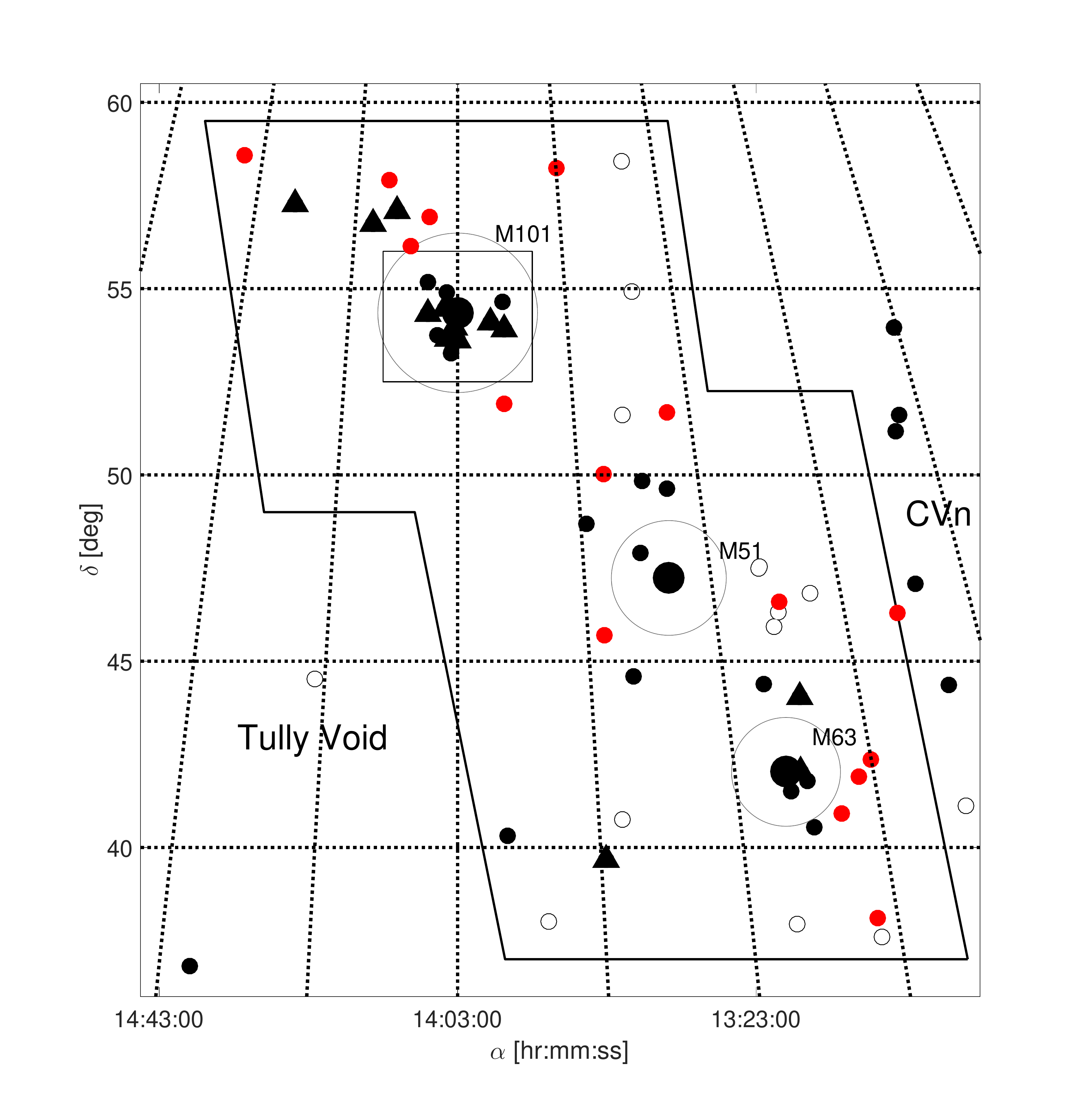}
 \caption{ Survey area of $\approx 330$ square degrees in the M101 group region. The black triangles are confirmed dwarf galaxies with distances in the M101 group complex. The small black dots are members based on their photometric properties, compiled from the Local Volume 
 Catalog \citep{2004AJ....127.2031K,2013AJ....145..101K}. The large black dots are the major galaxies that define the three subgroup centers in the region: 
 M\,101 (14h03m12.5s $+$54d20m56s), M\,51 (13h29m52.7s $+$47d11m43s), and M\,63 (13h15m49.3s $+$42d01m45s). The red dots indicate the positions of our 15 new dwarf candidates. Open circles are confirmed foreground (<5\,Mpc) galaxies taken from the LV Catalog. The footprint of the Dragonfly survey around M101 is outlined by the rectangle. The circles indicate the virial radii of $\approx 260$\,kpc {\citep[][for M\,101]{2014ApJ...787L..37M}} for the three major galaxies {(assuming the same virial radius for M\,51 and M\,63 as for M\,101)}.}
 \label{fieldImage}
\end{figure*}
The Sloan Digital Sky Survey \citep{2006AJ....131.2332G} contributed a tremendous amount to the knowledge of the dwarf galaxy population in the Local Group. Numerous resolved dwarfs were discovered by several teams \citep[e.g.][]{2010ApJ...712L.103B,2015ApJ...804L..44K}, pushing the limits of the known dwarfs into the regime of ultra-faint dwarf galaxies ($M_V > -4$ mag). There is currently no instrument which would enable us to discover such extremely faint galaxies outside of the LG. The typical limiting total luminosity reached beyond the LG is $M_V \sim -9.5$ mag, several magnitudes brighter than the LG ultra-faints. At the distance of M\,101, with SDSS data we can reach an absolute magnitude of $M_V \sim -10$.

Taking the LG Sculptor dwarf galaxy as a typical dwarf spheroidal galaxy (at the faint end of the dE luminosity function), we can assume $M_I=-4.1$\,mag and $V-I=1.5$\,mag for the tip magnitude and color of the Red Giant Branch (TRGB) \citep{2007MNRAS.380.1255R}. This translates into $M_r \approx -2.8$\,mag with $V-r \approx 0.2$\,mag.  Assuming a distance of 7\,Mpc for the dwarfs in the M\,101 group we calculate an apparent TRGB magnitude of $m_r=26.3$\,mag. The limiting magnitudes for point sources are provided by the SDSS Collaboration and are $g=23.3$\,mag and $r=23.1$\,mag, respectively \citep{2000AJ....120.1579Y}. Thus it becomes clear that in the SDSS data the TRGB is not resolved for dwarf galaxies in the M\,101 group and group complex.

For our search of unresolved dwarf galaxies we used 323 tiles in $g$ and $r$ from the SDSS Data Release 12 \citep{2015ApJS..219...12A}. Each tile covers a 1 square degree area put together in an online pipeline provided by the SDSS team, where the tiles overlap by 0.05$^{\circ}$ on each side, giving a gap-free survey area of $\approx$330 square degrees and covering the whole M\,101 group complex including the vicinities of M\,101, M\,51, and M\,63. Fig.\,\ref{fieldImage} shows the footprint of the survey. Also indicated is the much smaller footprint of the Dragonfly survey {\citep{2014ApJ...782L..24V,2014ApJ...787L..37M,2016ApJ...833..168M,2017arXiv170204727D}}.

\section{Search and detection of new dwarf candidates}
\begin{table}[ht]
\centering
\setlength{\tabcolsep}{3pt}
\caption{Names, coordinates, and morphological types of the 15 new dwarf galaxy candidates of the M\,101 group complex.}
\label{table:1}
\begin{tabular}{lccll}
\hline\hline
& $\alpha$ & $\delta$ & \\ 
Name & (J2000) & (J2000) & Type & Notes\\ 
\hline \\[-2mm]
M\,101 subgroup& & & & \\

dw1343+58 & 13:43:07 & $+$ 58:13:40 &BCD&  \\ 
dw1355+51 & 13:55:11 & $+$ 51:54:29 & dSph &   \\ 
dw1408+56 & 14:08:41 & $+$ 56:55:38 &dSph& \\ 

dw1412+56 & 14:12:11 & $+$ 56:08:31 &dSph&\\ 
dw1416+57 & 14:16:59 & $+$ 57:54:39 &dIrr/dSph& bg dwarf? \\
dw1446+58 & 14:46:60 & $+$ 58:34:04 &dSph& \\ 
\\
M\,51 subgroup& & & & \\
dw1313+46 & 13:13:02 & $+$ 46:36:08 & dIrr/BCD & bg spiral? \\

dw1327+51 & 13:27:01 & $+$ 51:41:08 &dSph& \\ 
dw1338+50 & 13:38:49 & $+$ 50:01:10 & dSph & bg dwarf? \\ 
dw1340+45 &13:40:37  & $+$ 45:41:54 &dIrr& \\ 
\\
M\,63 subgroup& & & & \\
dw1255+40 & 12:55:02 & $+$ 40:35:24 & dSph& CVn\,I mem? \\ 
dw1303+42 & 13:03:14 & $+$ 42:22:17 & dIrr&\\ 
dw1305+38 & 13:05:58 & $+$ 38:05:43 & dSph& bg dwarf?\\ 
dw1305+41 & 13:05:29  & $+$ 41:53:24 & dIrr/dSph&\\ 
dw1308+40 & 13:08:46  & $+$ 40:54:04  & dSph  &  \\  
%dw &  & $+$  &  &  \\ %  
%dw &  & $-$  &  \\ 
%dw &  & $-$  &  \\ 
\hline\hline
\end{tabular}
\end{table}

Lacking the power to resolve new faint dwarf galaxies into stars at that distance, we search for extended, low-surface brightness features. The surveyed region contains 29 known dwarf galaxies, {with 14 confirmed members via distance measurements, including the most recent Dragonfly dwarfs \citep{2017arXiv170204727D}, and 15} candidates where membership was estimated from their photometric and morphological properties. There are also 11 foreground dwarf galaxies known, with distance estimations smaller than 5\,Mpc. 

\begin{figure*}
%m101

\includegraphics[width=3.6cm]{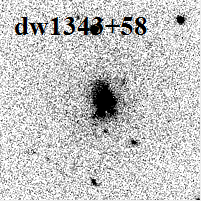}
\includegraphics[width=3.6cm]{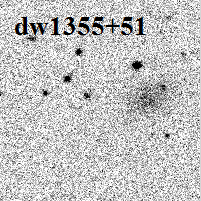}
\includegraphics[width=3.6cm]{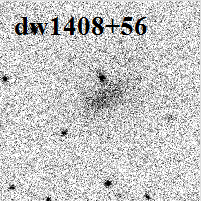}
\includegraphics[width=3.6cm]{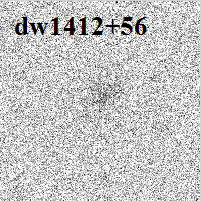}
\includegraphics[width=3.6cm]{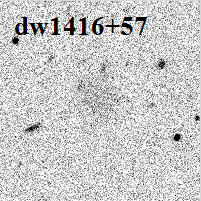}\\
\includegraphics[width=3.6cm]{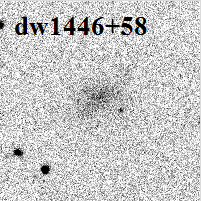}
%m51
\includegraphics[width=3.6cm]{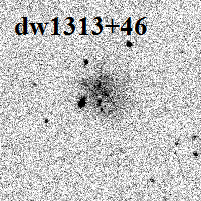}
\includegraphics[width=3.6cm]{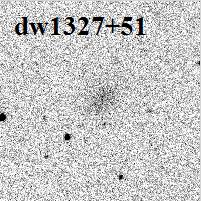}
\includegraphics[width=3.6cm]{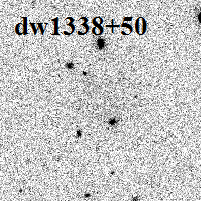}
\includegraphics[width=3.6cm]{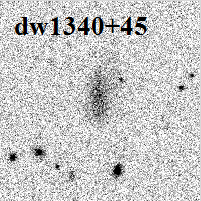}\\
%m63
\includegraphics[width=3.6cm]{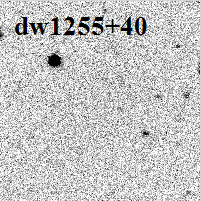}
\includegraphics[width=3.6cm]{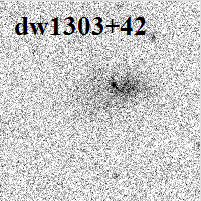}
\includegraphics[width=3.6cm]{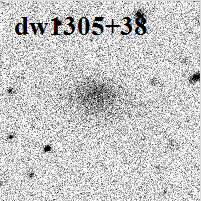}
\includegraphics[width=3.6cm]{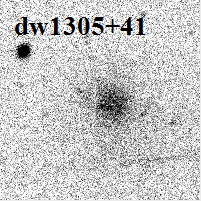}
\includegraphics[width=3.6cm]{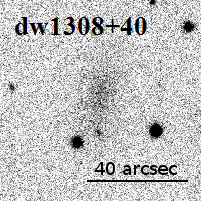}
\caption{Gallery showing SDSS $r$-band images of the 15 new M\,101 group dwarf galaxy candidates. Note that dw1355+51 is at the edge of the CCD and is not centered in the stamp but offset to the right, and dw1255+40 is barely visible without a strong Gauss convolution.
One side of an image corresponds to 80\,arcsec or 2.7\,kpc at the distance of 7\,Mpc. North is to the right, east to the top. }
\label{sample0}
\end{figure*}

Each tile was first binned (mapping $9\times 9$ pixels onto 1 pixel using the mean value) and convolved with a $3\times 3$ pixel Gauss kernel. This dramatically increased the signal-to-noise ratio by a factor of $\sim30$ and thus the visibility of low-surface brightness features against the background sky. The tiles were visually inspected by two people from our team (OM and RS), where the gray scale was varied such that different dynamical ranges could be examined.
This procedure led to the discovery of 15 new dwarf galaxy candidates in the M\,101 group complex (red dots in Fig.\,1). Their coordinates and morphological classification are compiled in Table\,\ref{table:1} and the candidate images presented in Fig.\,\ref{sample0}. {We classified the candidates according to their morphological appearance: objects which appear symmetric, diffuse and elliptical as dSph (dwarf spheroidal); objects with an uneven brightness distribution, e.g. due to HII regions, as dIrr (dwarf irregular); and objects with a clumpy, high-surface brightness central component and a diffuse halo as BCD (Blue Compact Dwarfs). There are three cases where the morphology is ambiguous. We present the two possible classes separated with a slash, e.g. dIrr/dSph, where the first is the more likely morphological type.}

Note that we have assigned each dwarf galaxy candidate to one of the three subgroups in Table\,\ref{table:1}. The assignment is based on the shortest angular distance to either M\,101, M\,51, or M\,63. We will use the individual parent galaxy's distance to calculate absolute magnitudes for the new candidates as distances of the three major galaxies systematically differ.

The biggest challenge of our Cen\,A survey \citep{2015A&A...583A..79M,2017A&A...597A...7M} was the contamination of foreground stars and Galactic nebulae (cirrus). Cirrus can appear in every shape and size and thus can mimic the morphology of a dwarf galaxy. Fortunately, our survey area around M\,101 is at high Galactic latitudes ($b\approx 60^\circ$), i.e. far away from the Galactic plane where the density of Galactic cirrus and foreground stars is supposedly very small. Hence the problem of false positive detections is minimal.

\begin{figure}[ht]
\centering
  \includegraphics[width=9cm]{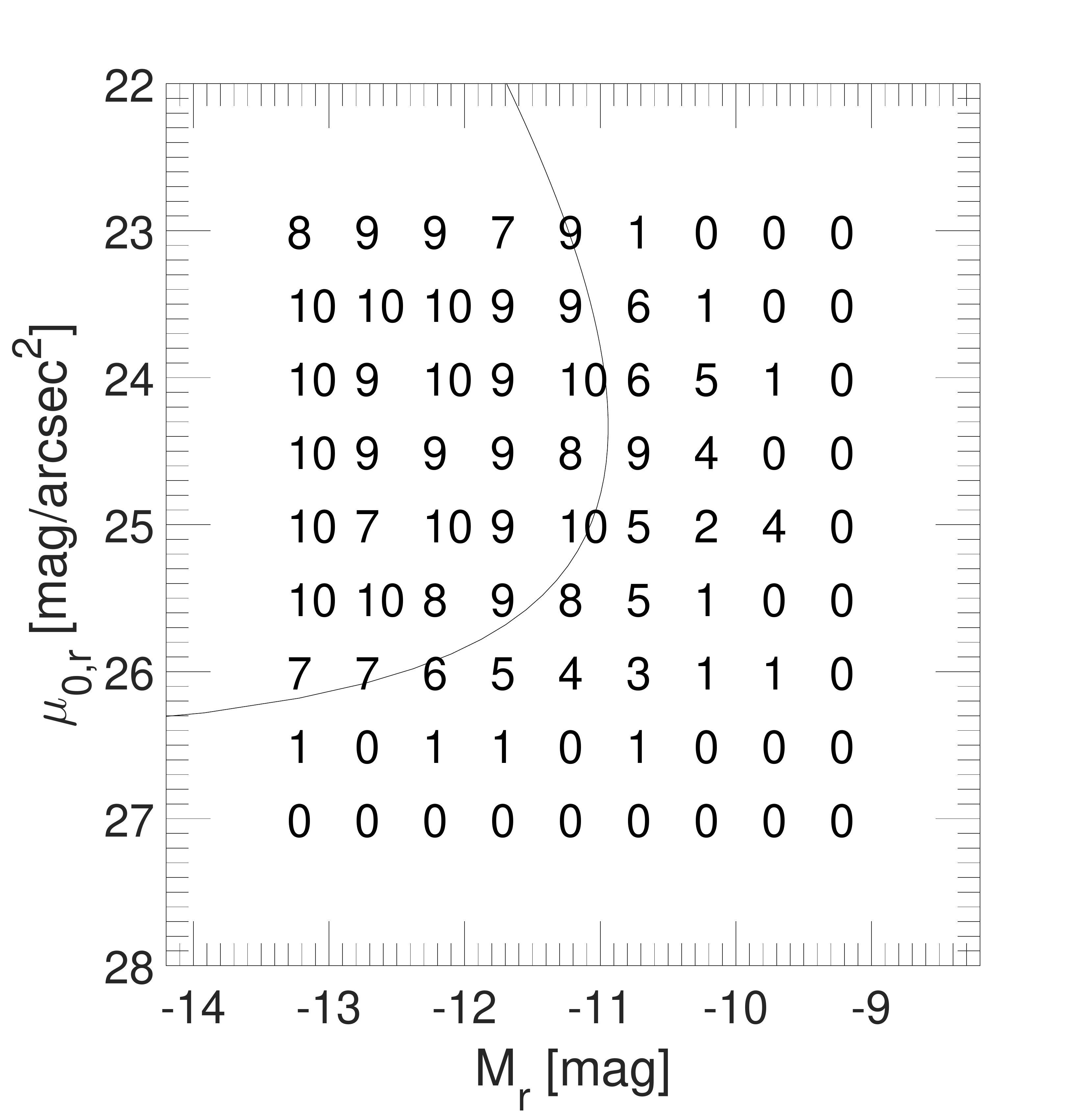}
\caption{Results of our artificial galaxy detection test shown as an array of numbers indicating the detection efficiency in the surface brightness-absolute magnitude plane. A 10 means 100\,\% detection, a 0 no detection. The test array is divided into half-magnitude bins. The thin line corresponds to the estimated {70\%} completeness boundary, see text for the formula. }
\label{art}
\end{figure}

Nevertheless, we performed artificial galaxy tests to estimate our detection efficiency and the depth (surface brightness limit) of the survey. For this we superimposed artificial galaxies on real images in two different tiles. {The profiles for the artificial galaxies were created using a S\'ersic profile with $n=1$ (exponential profile, see below for the formula)}. The central surface brightness range was between 23 and 27\,mag arcsec$^2$ and the apparent magnitude range between 16 and 20\,mag, with a step size of 0.5\,mag and 0.5\,mag arcsec$^{-2}$, respectively. At a distance of 6.95\,Mpc this corresponds to absolute magnitudes between $-13.2$ and $-9.2$. This gives a total of 49 galaxies in an array of the surface brightness-absolute magnitude plane to detect per mosaic and iteration. We did five iterations where in each iteration we randomly placed all artificial galaxies into a $r$ band tile. This was repeated for two different tiles such that we had ten iterations in total. In Fig.\,\ref{art} the results are presented in a $\mu_{0,r} - M_r$ diagram. The number of times an artificial galaxy was detected is plotted, zero means no detection and ten corresponds to a 100\,\% detection rate. We do not expect a detection rate of 100\,\%  {even for well detectable galaxies} because artificial galaxies can be randomly placed behind bright and extended stars or galaxies. Important to note is that not all parameter combinations lead to reasonable LSB dwarf galaxies, e.g. a high $\mu_{0,r}$ value (low SB) together with a small $M_r$ value (relatively high luminosity) will lead to very extended and faint objects, {which are not found in the Local Group, see Fig.\,\ref{rel}}. {There is no significant difference in the detection rate between the two different tiles.}

It can be seen that essentially all artificial objects with $\mu_{0,r} \leq$ 25.5 $r$ arcsec$^{-2}$ and $M_r \leq -11$ are detected. A more appropriate completeness boundary of the survey is provided by the following analytic forms \citep[see][]{1990PhDT.........1F, 1988AJ.....96.1520F}:
$$m_{tot}= \mu_0 - 5\log{(r_{lim})} - 2.18 + 5\log{(\mu_{lim}-\mu_0)}$$
$$m_{tot}= \mu_{lim} - \frac{r_{lim}}{0.5487 r_{eff}}- 2.5 \log{(2\pi(0.5958 r_{eff})^2)}$$
where all galaxies larger than $2 r_{lim}$ within a given isophotal level of $\mu_{lim}$ should be detected. The first equation is for the $\mu_{0,r} - M_r$ relation, the second one is for the $r_{eff}-M_r$ relation, {where $r_{eff}$ corresponds to the half-light radius of the object}. To calculate the absolute magnitude $M_r$ we assumed a distance of 7\,Mpc. We estimated the two parameters such that the completeness boundary would contain all bins of the $\mu-M$ array where the detection rate is higher than 70\,\%, resulting in $r_{lim}=13\arcsec$ and $\mu_{lim,r}=26.4$\,$r$ mag arcsec$^{-2}$. This completeness curve is shown in Fig.\,\ref{art}.

The {identification of high-surface brightness dwarf members (against a background of apparently small spiral galaxies) is more difficult and our detection efficiency for these objects} cannot easily be assessed by an artificial galaxy test. In general we have to expect that potential high SB dwarf members of the M\,101 group complex essentially go unnoticed in our survey. However, one rather convincing case of a blue compact dwarf (BCD, dw1343+58) has been found. {On the other hand,} a good low SB candidate could of course be in the near background (hence the remark `bg dwarf?' in Table 1, where this seemed equally possible) or the near foreground. In fact, towards the western boundary, the M\,101 group region overlaps in the sky with the closer Canes Venatici (CVn) cloud (see also Fig.\,1). One candidate in the M\,63 subgroup, dw1255+40, is indeed a possible member of that cloud. The problem of confusion is more generally addressed in Sect.\,6.1 below.  

We rejected candidates that were close to ultra violet sources (UvS, e.g. brilliant young foreground stars) listed in the Nasa Extragalactic Database. Such sources can illuminate surrounding dust clouds and make them appear as faint low-surface brightness objects. While closely resembling the morphology of a dwarf spheroidal galaxy the unusually blue color ($g-r<0.1$\,mag) of such objects raises doubts for these objects to be dwarf candidates. A good example was found at the coordinates 14h09m12s, +51d13m27s, which is only 0$\arcmin$.136 separated from an UvS. It mimics the morphology and has structural parameters of a diffuse dwarf galaxy, but was suspiciously blue.

The reader may wonder why we did not use the SDSS data reduction pipeline directly for the detection of dwarf-like low surface brightness objects. There is indeed a tool implemented in the data reduction pipeline for the detection of extended sources. However, \citet{2004AJ....127..704K} pointed out that galaxies are shredded by this tool, as different luminosity knots from the same source are detected and defined as separate, individual SDSS objects. Tests have shown \citep{2004AJ....127..704K} that the SDSS pipeline tool is unsuited for the detection of LSB objects, giving only a low detection rate of test galaxies and too many false detections. Nevertheless, we checked the SDSS database for the presence of any kind of detection counterpart for our new candidates. Indeed, all our candidates have matches in the SDSS database, but the link between these SDSS objects and a possible group membership of M\,101 was not made before the present work.
In addition, the SDSS photometry for these low-surface brightness objects is unreliable {as stated by the SDSS photometry pipeline for those objects}. The SDSS database also provides redshifts when available, but none of our candidate galaxies, not even the high surface brightness dw1343+58, has a measured redshift. 

As alluded to in the introduction, it is interesting to note that six out of seven candidate members from the Dragonfly survey are clearly visible in the SDSS images{, thus were re-detected in our survey}, which strongly suggests that the SDSS data still contains a lot of hidden treasures waiting to be discovered. This is insofar not surprising as the central surface brightness range of those candidates is between 25.1 and 26.8 $r$ mag arcsec$^{-2}$, which is still detectable according to our artificial galaxy tests.

\section{Surface photometry}
\begin{figure*}
\includegraphics[width=3.6cm]{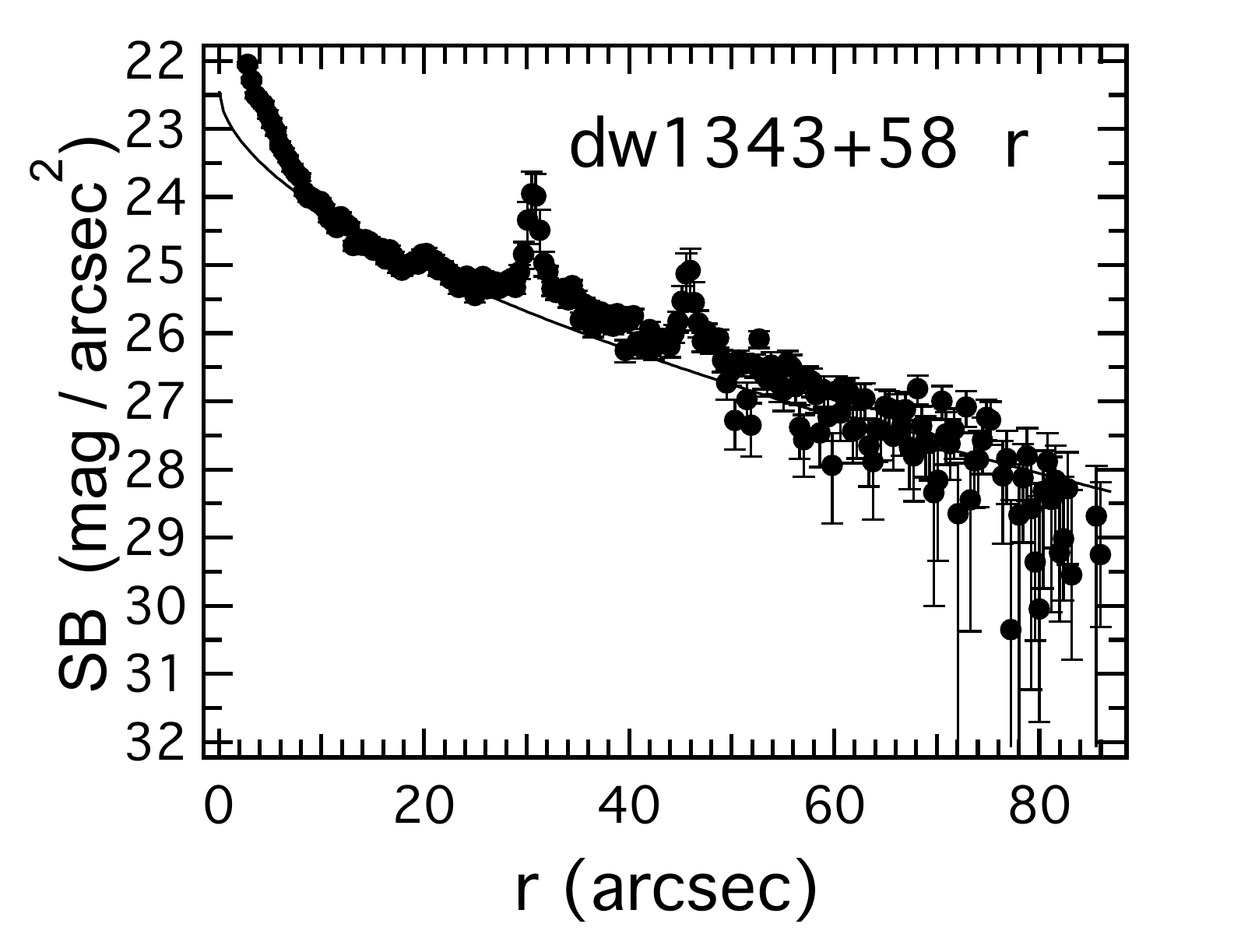}
\includegraphics[width=3.6cm]{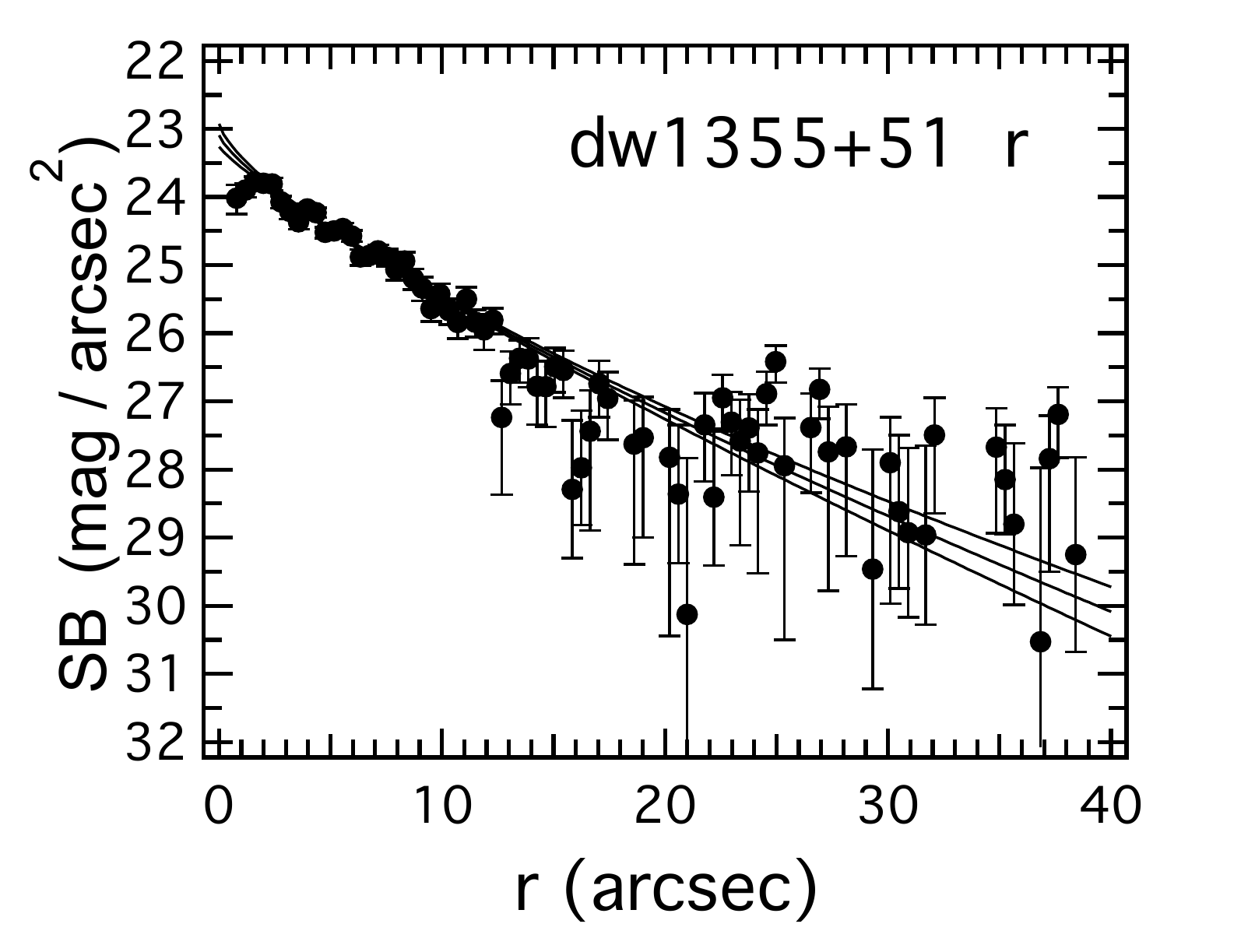}
\includegraphics[width=3.6cm]{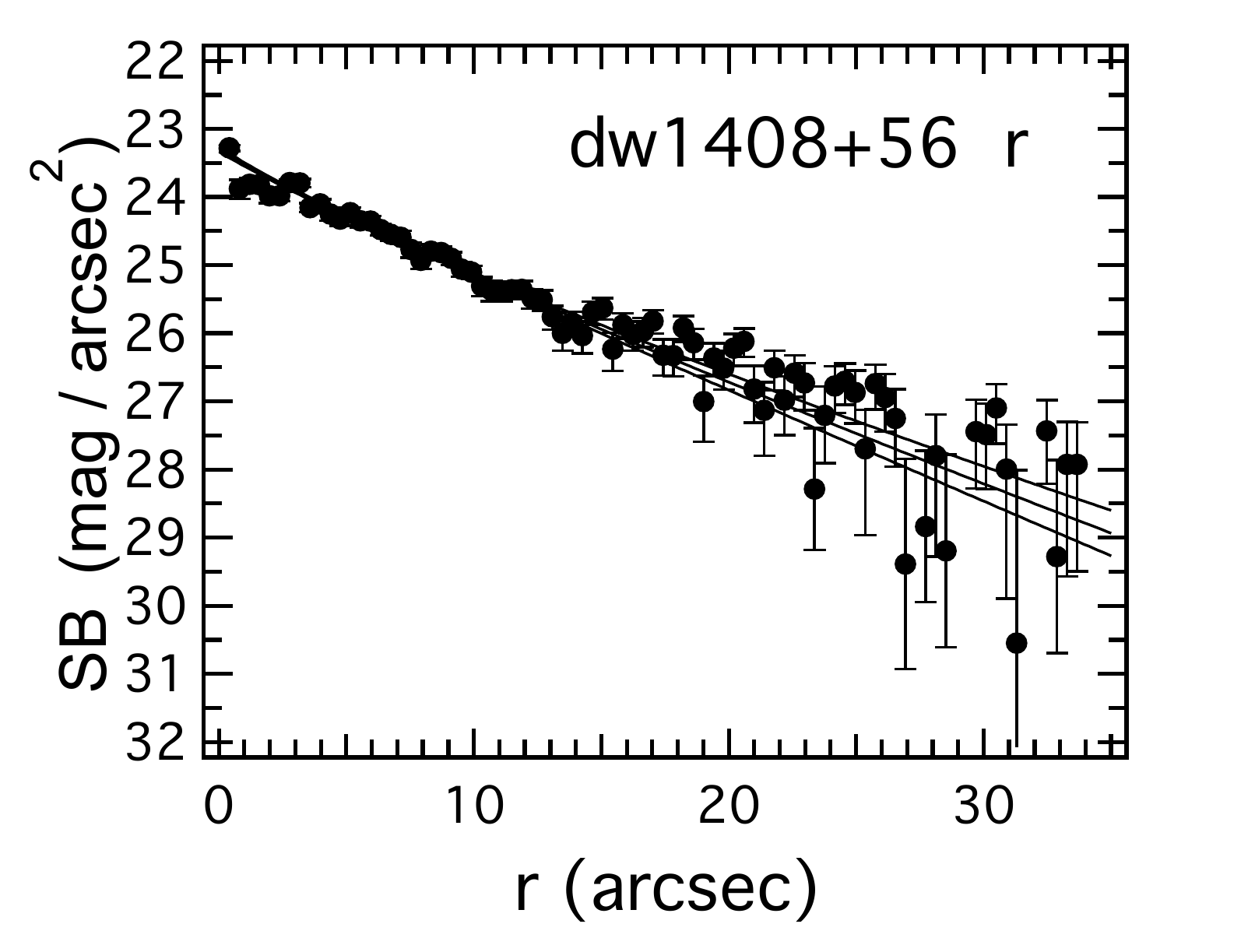}
\includegraphics[width=3.6cm]{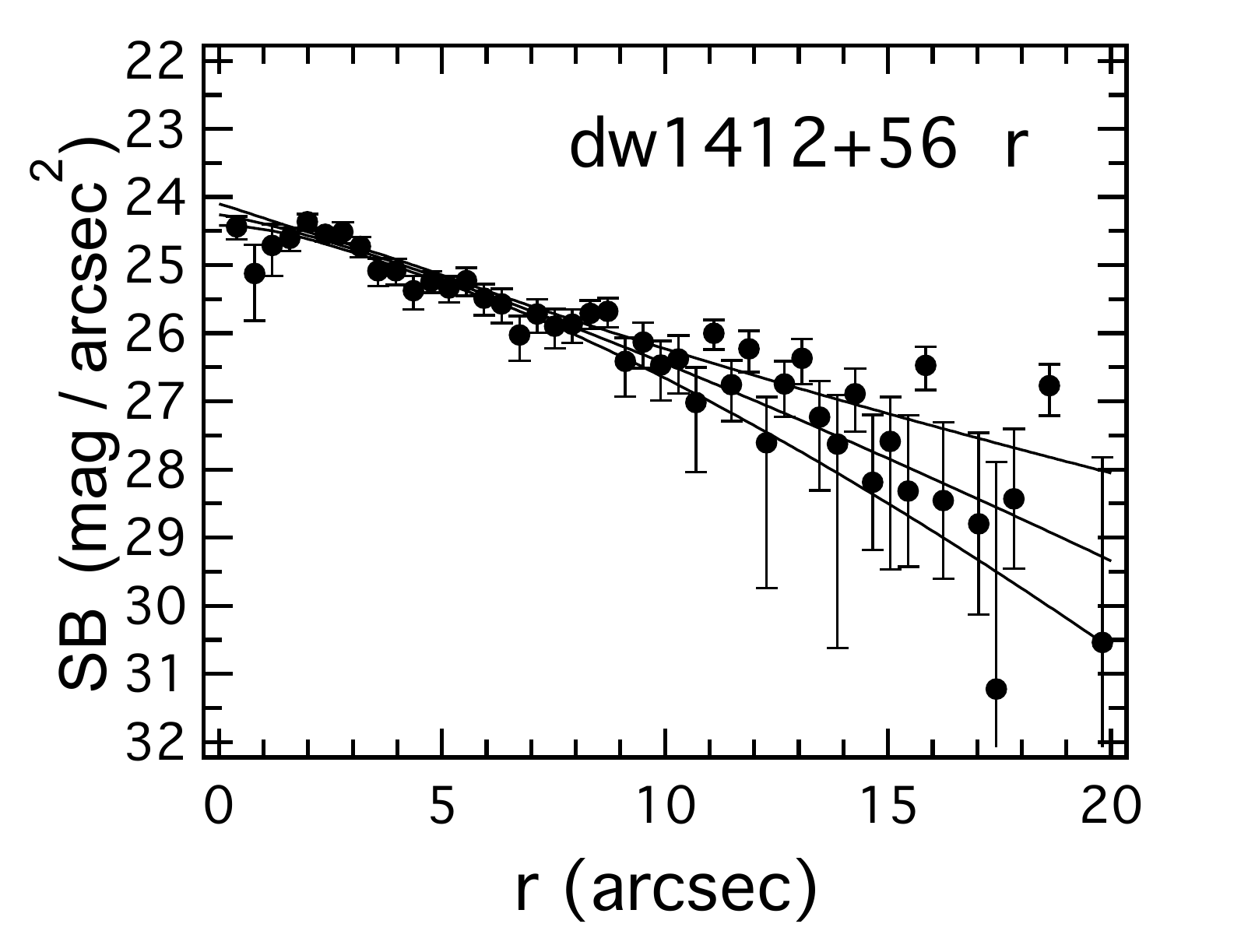}
\includegraphics[width=3.6cm]{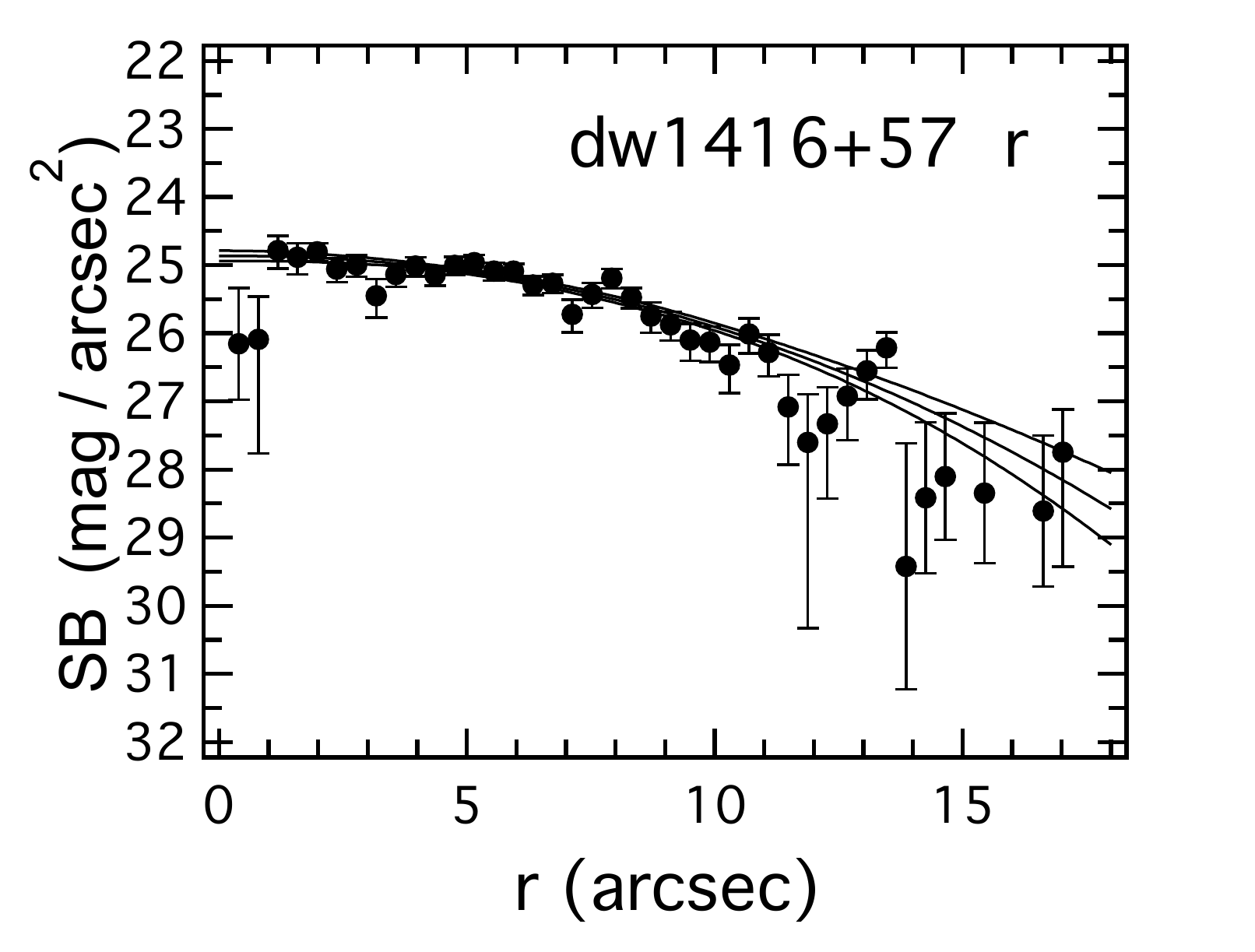}\\
\includegraphics[width=3.6cm]{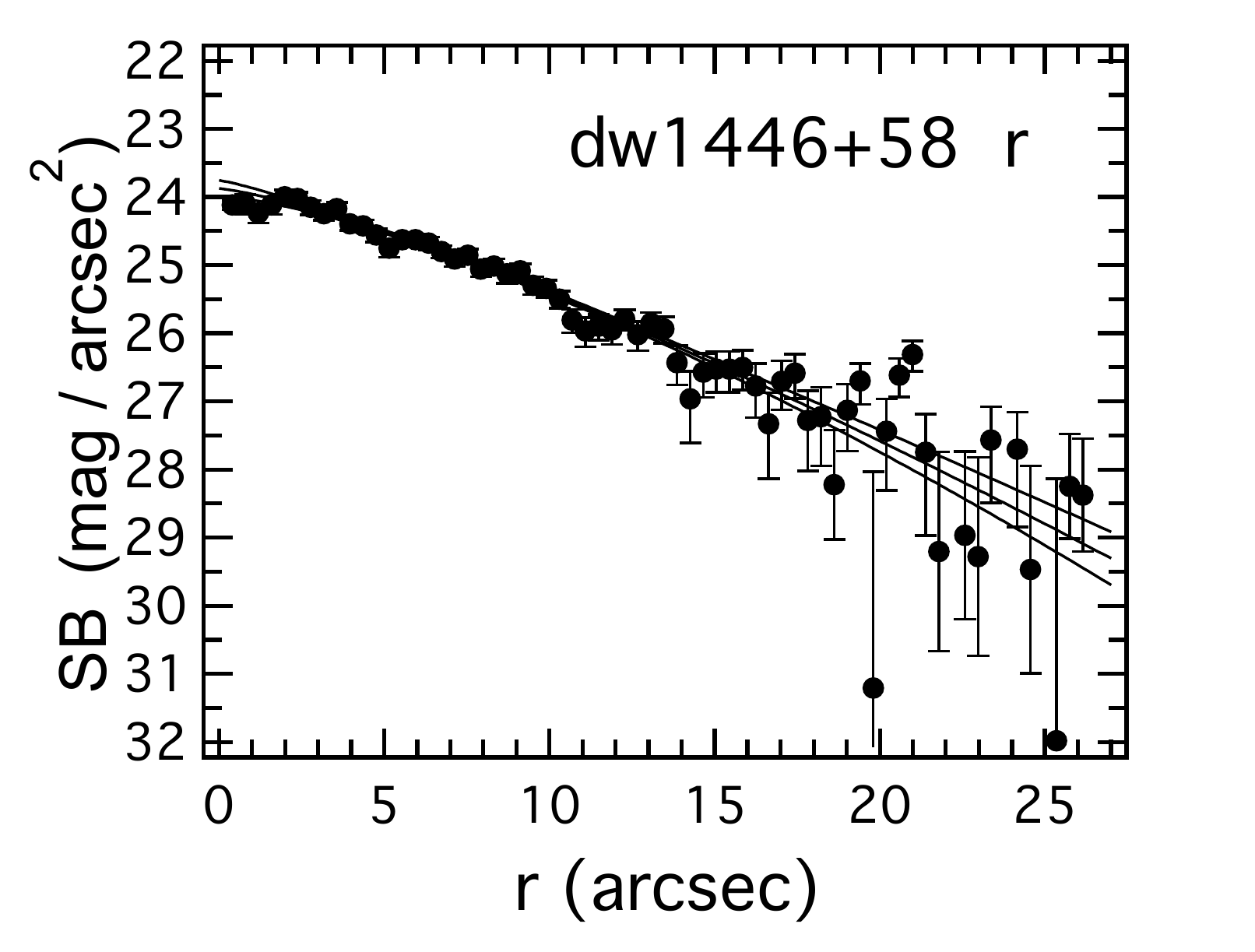}
%m51
\includegraphics[width=3.6cm]{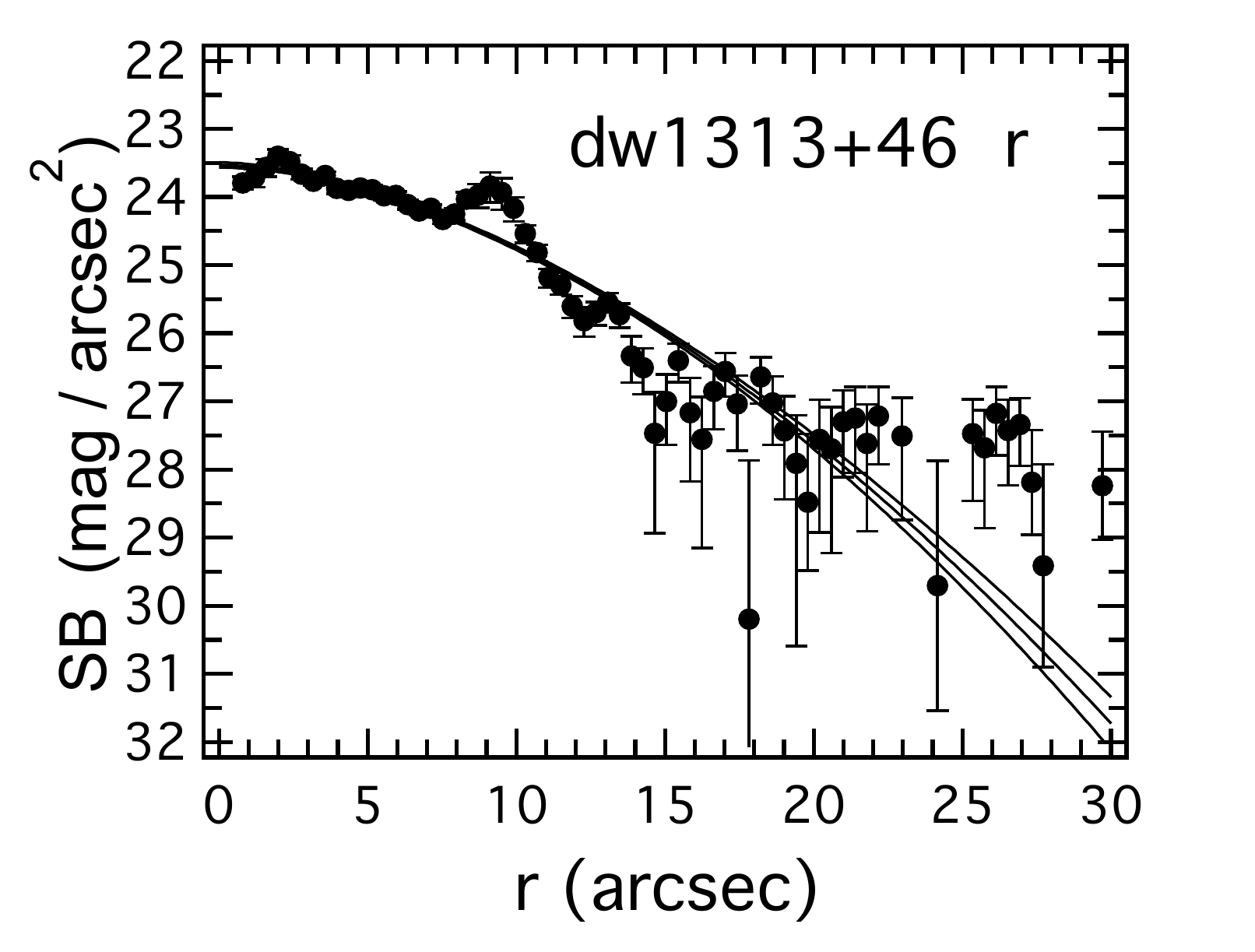}
\includegraphics[width=3.6cm]{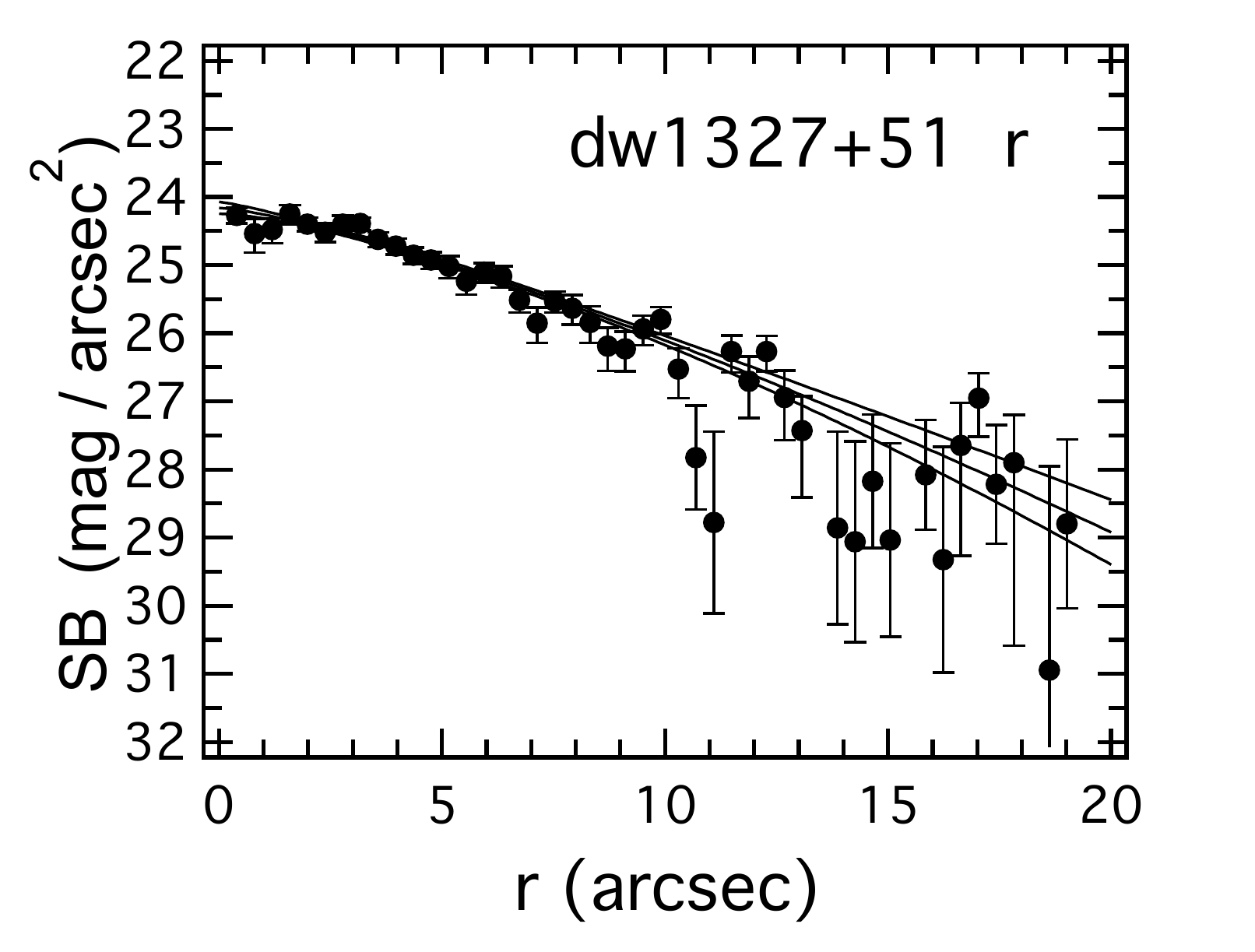}
\includegraphics[width=3.6cm]{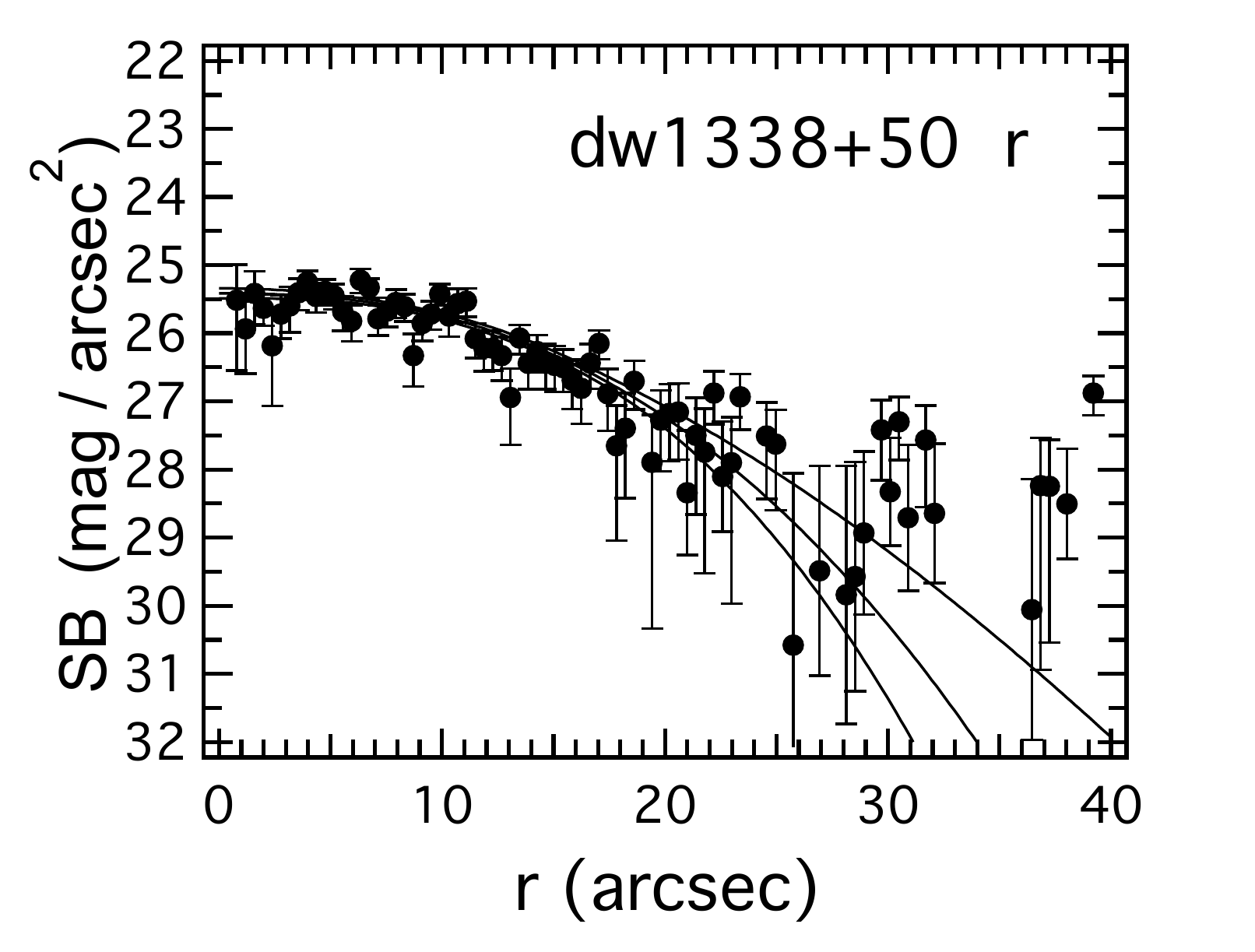}
\includegraphics[width=3.6cm]{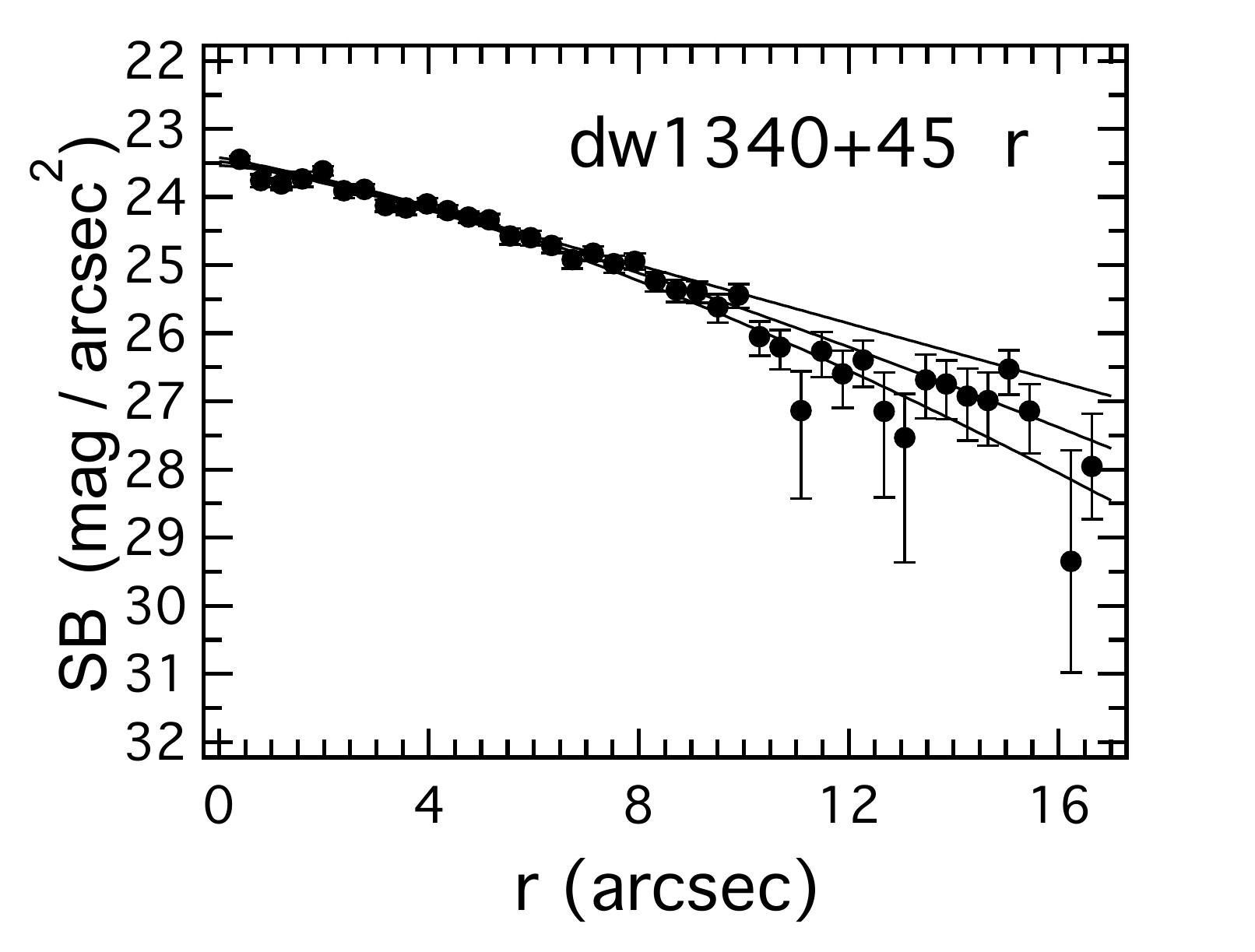}\\
%m63
\includegraphics[width=3.6cm]{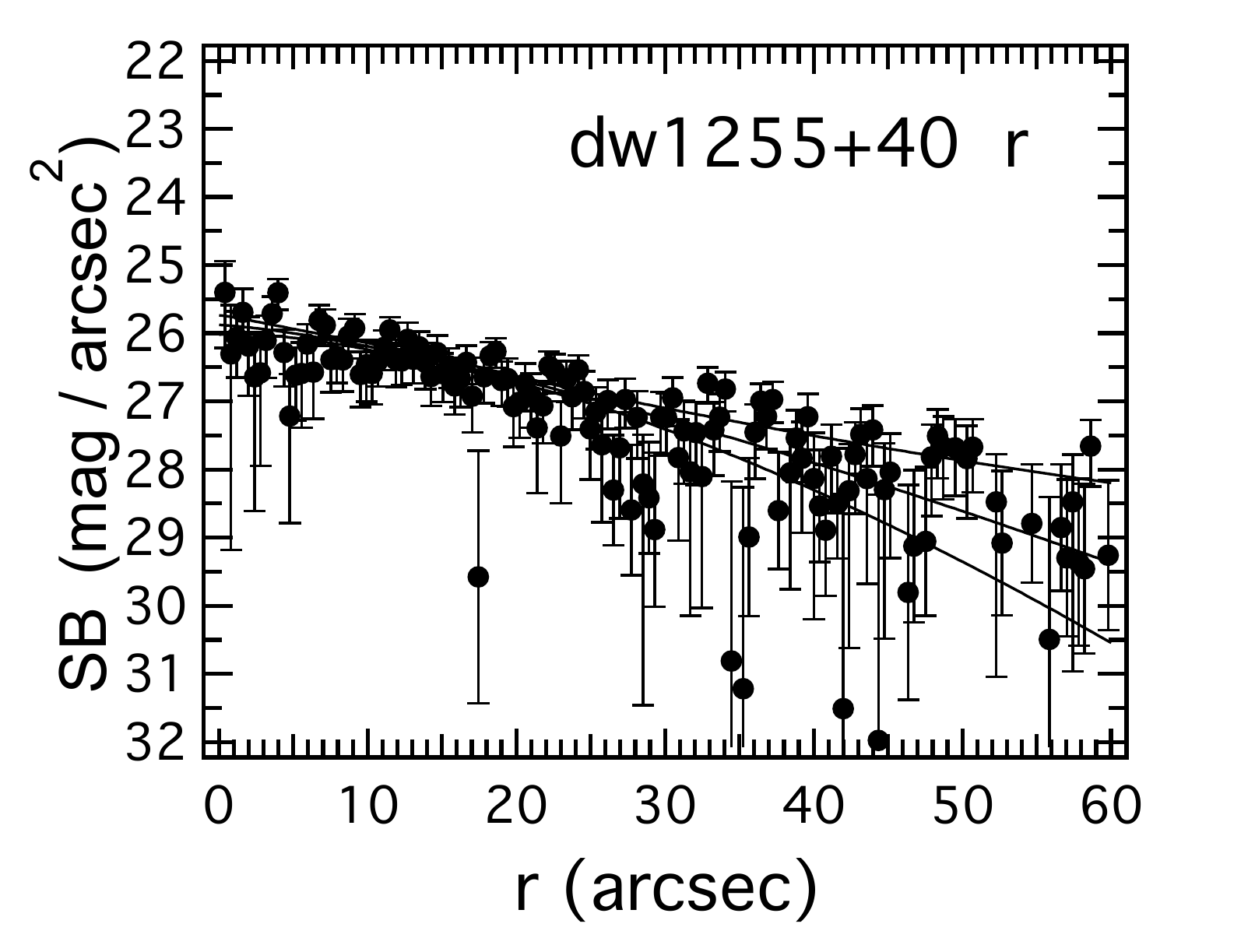}
\includegraphics[width=3.6cm]{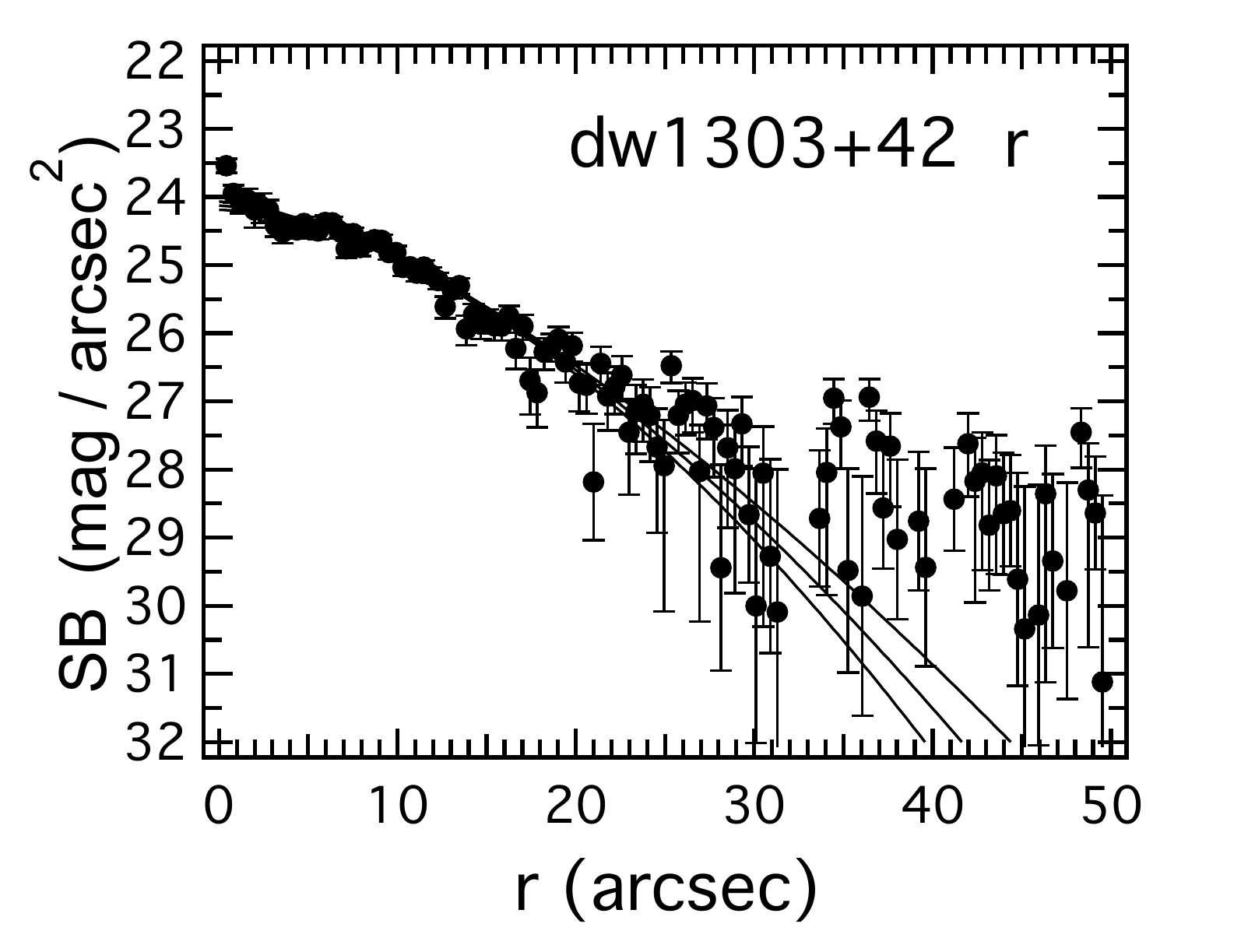}
\includegraphics[width=3.6cm]{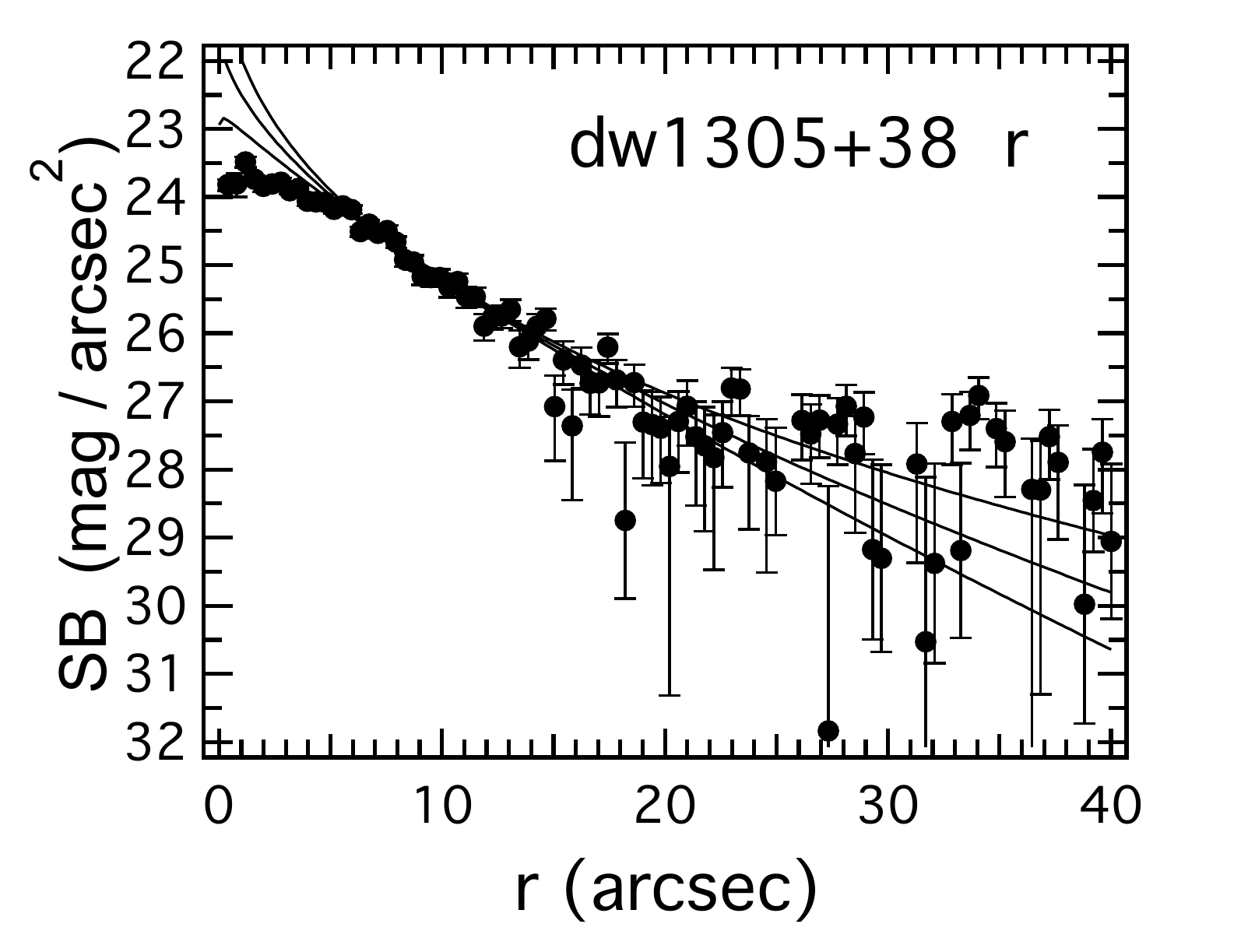}
\includegraphics[width=3.6cm]{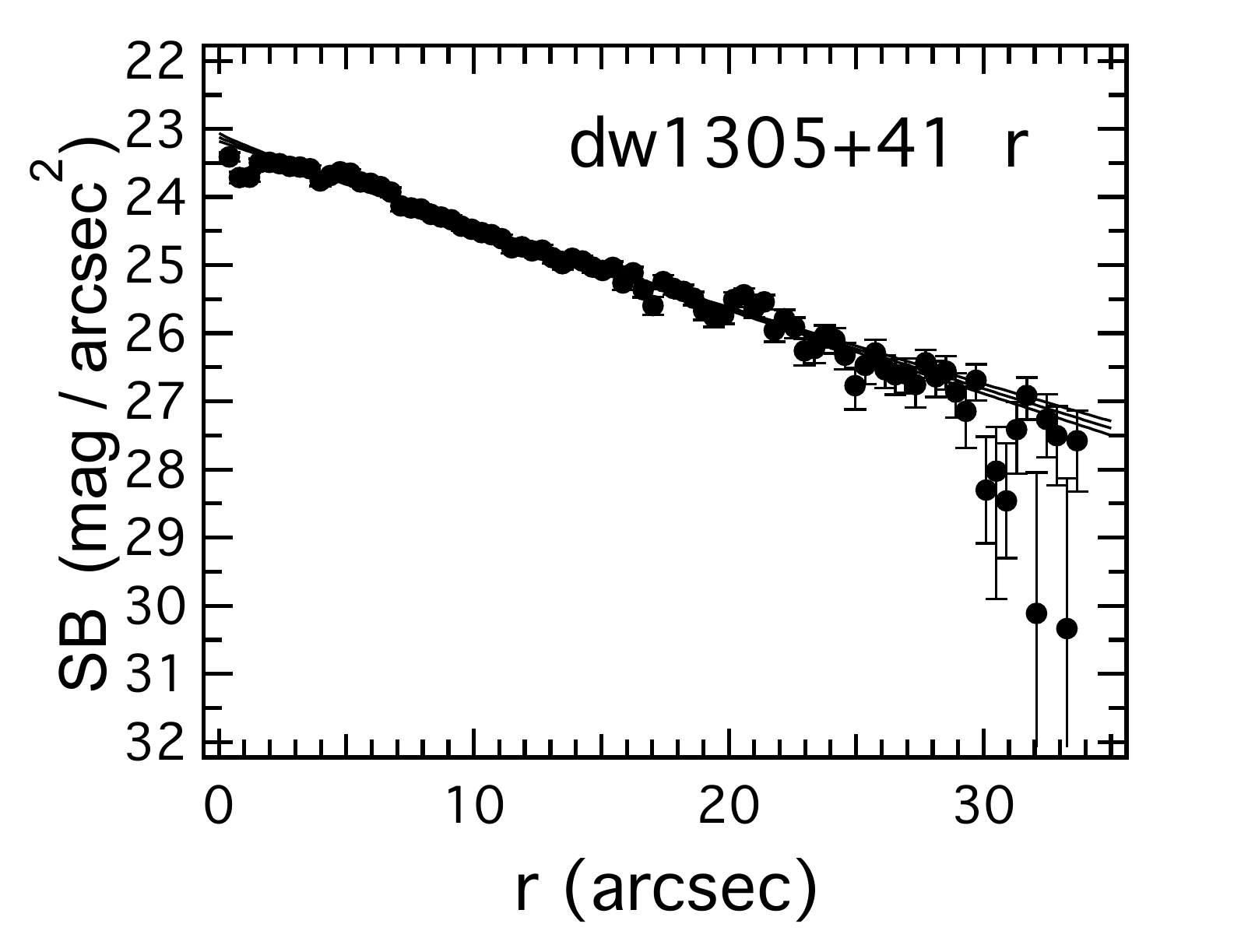}
\includegraphics[width=3.6cm]{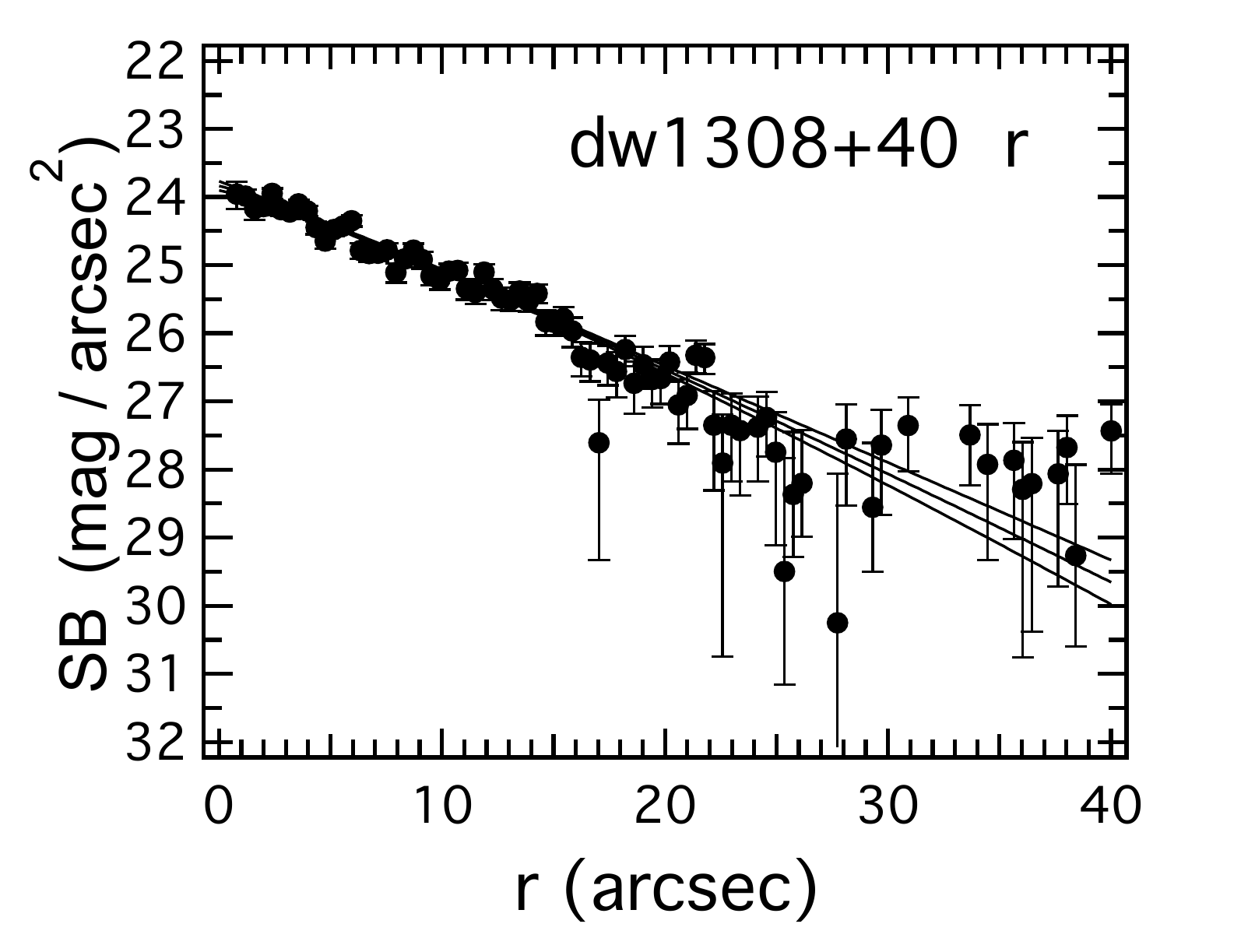}

\caption{Surface brightness profiles of all dwarf galaxy candidates in $r$ and the best-fitting S\'ersic profiles with $1 \sigma$ confidence intervals.}
\label{sbp}
\end{figure*}

We performed $gr$ surface photometry for the new candidates in the surveyed area. Cosmic rays, foreground stars and background galaxies were replaced with patches of sky from the surrounding area using IRAF to maintain the statistical properties of the local sky background. The nominal galaxy center was determined using a circle that best represents the shape of the outer isophotes of the galaxy. We \text{emphasize} that this center is a proxi for the underlying mass distribution, but does not necessarily coincide with the location of maximum surface brightness.  The sky background was estimated from varying the galaxy growth curve until it became asymptotically flat. We computed for each galaxy its total apparent magnitude, the mean effective surface 
brightness $\langle \mu\rangle_{eff} $, and the effective radius $r_{eff}$ in both bands. We used a circular aperture to measure the surface brightness profiles with a step size of $0\farcs396$ (corresponding to 1 pixel).
S\'ersic profiles  \citep{1968adga.book.....S} were fitted at the radial surface brightness profiles using the equation:
$$\mu_{sersic}(r)= \mu_0+1.0857\cdot\left(\frac{r}{r_0}\right)^{n}$$
where $\mu_0$ is the S\'ersic central surface brightness, $r_0$ the S\'ersic scale length, and $n$ the S\'ersic curvature index. See Fig.\,\ref{sbp} for all surface brightness profiles in the $r$ band and the associated S\'ersic fits.

The total magnitude uncertainty is estimated to be around $\approx$\,0.3\,mag. It is made up from uncertainties related to foreground star removal ($\approx$\,0.2\,mag) and sky background estimation ($\approx$\,0.2\,mag). {The uncertainty for the mean effective surface brightness is driven by the uncertainty in the measured total apparent magnitude. The error for the half-light radius ($\approx$\,1.3\,arcsec) is given by the determination of the growth curve. See \citet{2015A&A...583A..79M} for a more detailed explanation of the uncertainty estimates}. Numerical uncertainties for the S\'ersic parameters are shown in Table\,\ref{table2}.

\begin{figure}[ht]
\centering
  \includegraphics[width=9cm]{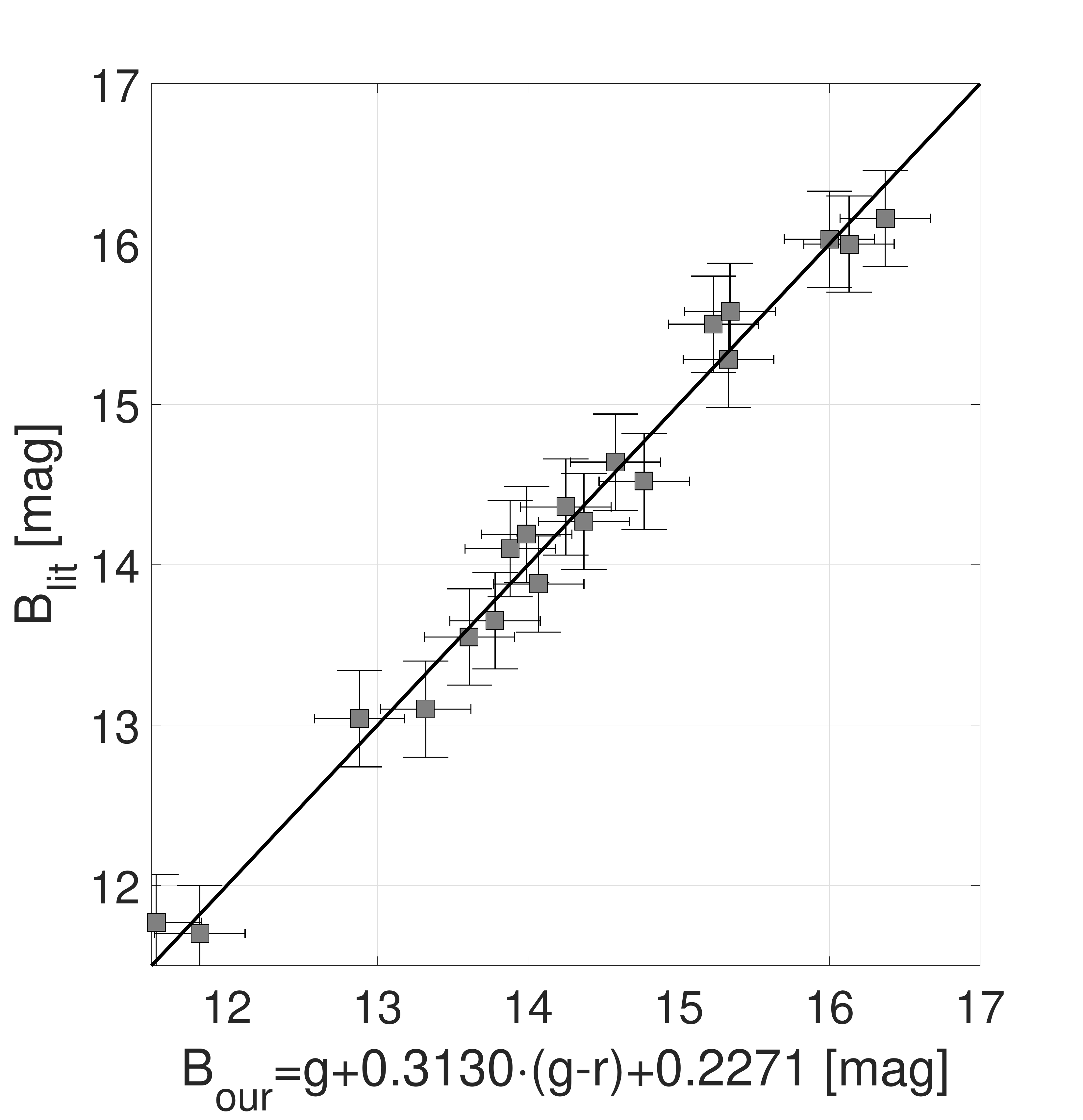}
\caption{Our photometry for 19 known M\,101 dwarf galaxies converted to $B$ band versus the literature values taken from \citet{1999A&AS..137..337B}. The line corresponds to unity.}
\label{diff}
\end{figure}

To compare our $gr$ photometry and the structural parameters with dwarf galaxies in the literature, we used the following transformation equations \citep{SloanConv}:
$$
V = g - 0.5784\cdot(g - r) - 0.0038 
$$
$$
B = r + 1.3130\cdot(g - r) + 0.2271
$$
In \citet{2017A&A...597A...7M} we tested the quality of our photometry against literature values. The agreement was well within the uncertainties. In the same spirit we conducted a comparison of the photometric values for 19 known dwarfs in the field of M\,101 taken from \citet{1999A&AS..137..337B} with our own SDSS photometry. The values are in excellent agreement within our error estimates (see Fig.\,\ref{diff}). We measure a standard deviation of $\sigma_{\Delta B} =0.18$\,mag and a mean of $\mu_{\Delta B}=0.00$\,mag was calculated. 

As stated earlier, the M101 survey area is at high Galactic latitudes, therefore the Galactic extinction values for the $g$ and $r$ band are less than 0.05\,mag,
much smaller than the photometric uncertainties. Hence, no corrections for Galactic extinction were applied when  calculating absolute magnitudes.

In Table\,\ref{table2} we present the photometric data for the 15 newly detected dwarf galaxy candidates in the M\,101 group complex. 
The quantities listed are as follows:
(1) name of candidate;
(2+3) total apparent magnitude in the $g$ and $r$ bands;
(4) absolute $r$ band magnitude. For candidates of the M\,101 subgroup  
the mean distance of the M\,101 group (6.95\,Mpc) is assumed, for candidates of the M\,51 and M\,63 subgroups the distances of these major galaxies, 8.6 and 9.0\,Mpc, respectively;
(5) integrated $g-r$ color;
(6) S\'ersic central surface brightness in the $r$ band;
(7) S\'ersic scale length in the $r$ band;
(8) S\'ersic curvature index in the $r$ band;
(9) mean effective surface brightness in the $r$ band;
(10) effective radius in the $r$ band.
These photometric data will be used in the discussion section 6.1 below to assess the M\,101 group complex membership of the candidates. 
(11) the logarithm of the effective radius in the $r$ band, converted to pc with a distance assumption according to the subgroup.
\begin{table*}

\caption{Photometric and structural parameters of the new dwarf candidates in the surveyed region of the M\,101 group complex.}
\centering
\small
\setlength{\tabcolsep}{3pt}
\begin{tabular}{lccrrcrccrr}
\hline\hline 
Name & ${g_{tot}}$ & ${r_{tot}}$  & $M_{r}$ & $(g-r)_{0,tot}$ & $\mu_{0,r}$ & $r_{0,r}$ & $n_r$ & $\langle\mu\rangle_{eff,r}$ &  $r_{eff,r}$ & $\log r_{eff,r}$\\ 
 & mag & mag & mag & mag &  mag arcsec$^{-2}$ &arcsec & & mag arcsec$^{-2}$ &arcsec & $\log$ pc \\ 
 (1)& (2) & (3) &  (4) & (5) &  (6) & (7) &(8) & (9) & (10) & (11) \\ 
\hline \\
M\,101 subgroup & & & & & & & & &\\
dw1343+58 & 15.54 & 15.17 & -14.0 & 0.370 & $18.93 \pm 2.26 $  & $0.04 \pm 0.77 $  & $ 0.27 \pm 0.12 $  & 24.45 & 28.6 & 2.98\\
dw1355+51 & 18.76 & 18.09 & -11.1 & 0.666 & $23.09 \pm 0.17 $  & $3.67 \pm 0.69 $  & $0.78 \pm 0.07 $  & 24.44 & 7.44 & 2.39\\
dw1408+56 & 18.01 & 17.50 & -11.7 & 0.507 & $23.28 \pm 0.06 $  & $5.48 \pm 0.38 $  & $0.89 \pm 0.06 $  & 24.71 & 11.0 & 2.57\\
dw1412+56 & 19.46 & 18.75 & -10.5 & 0.702 & $24.26 \pm 0.16 $  & $5.64 \pm 0.80 $  & $1.22 \pm 0.29 $  & 25.29 & 8.08 & 2.43\\
dw1416+57 & 19.06 & 18.83 & -10.4 & 0.227 & $24.86 \pm 0.08 $  & $10.19 \pm 0.47 $  & $2.16 \pm 0.31 $  & 25.16 & 7.35 & 2.39\\
dw1446+58 & 18.46 & 17.90 & -11.3 & 0.559 & $23.87 \pm 0.12 $  & $7.62 \pm 0.74 $  & $1.27 \pm 0.13 $  & 24.66 & 8.97 & 2.48\\
 \\
 M\,51 subgroup & & & & & & & & &\\
 dw1313+46 & 17.63 & 17.36 & -12.3 & 0.274 & $23.53 \pm 0.04 $  & $9.33 \pm 0.26 $  & $1.73 \pm 0.07 $  & 23.78 & 7.69 & 2.50\\
 dw1327+51 & 19.34 & 18.79 & -10.9 & 0.550 & $24.16 \pm 0.09 $  & $6.36 \pm 0.51 $  & $1.29 \pm 0.15 $  & 24.74 & 6.19 & 2.41\\
 dw1338+50 & 19.15 & 18.35 & -11.3 & 0.809 & $25.41 \pm 0.07 $  & $16.16 \pm 0.66 $  & $2.43 \pm 0.42 $  & 25.62 & 11.3 & 2.67\\
 dw1340+45 & 18.28 & 18.14 & -11.5 & 0.136 & $23.48 \pm 0.06 $  & $5.76 \pm 0.29 $  & $1.25 \pm 0.18 $  & 24.24 & 6.61 & 2.44\\
 \\
 M\,63 subgroup & & & & & & & & &\\
dw1255+40 & 18.41 & 17.82 & -11.9 & 0.594 & $25.88 \pm 0.14 $  & $25.14 \pm 2.48 $  & $1.34 \pm 0.42 $  & 26.49 & 21.6 & 2.97\\
dw1303+42 & 18.06 & 17.29 & -12.5 & 0.770 & $24.13 \pm 0.06 $  & $12.19 \pm 0.54 $  & $1.61 \pm 0.12 $  & 24.56 & 11.3 & 2.69\\
dw1305+38 & 17.69 & 17.51 & -12.3 & 0.178 & $21.57 \pm 1.37 $  & $1.31 \pm 1.87 $  & $0.59 \pm 0.23 $  & 24.41 & 9.57 & 2.62\\
dw1305+41 & 17.06 & 16.70 & -13.1 & 0.354 & $23.12 \pm 0.06 $  & $8.21 \pm 0.53 $  & $0.94 \pm 0.05 $  & 24.14 & 12.2 & 2.72\\
dw1308+40 & 18.19 & 17.54 & -12.2 & 0.650 & $23.84 \pm 0.07 $  & $8.81 \pm 0.60 $  & $1.11 \pm 0.07 $  & 24.74 & 11.0 & 2.68\\
\hline\hline
\end{tabular}
\label{table2}
\end{table*}

\section{Geometrical alignment}

In preparation for an analysis of the spatial structure of the M\,101 group complex (Sect.\,6.2 below) we first define a natural spatial reference frame for the complex by fitting a plane through the galaxy positions in the close environment of M\,101 itself. In a similar manner
\cite{2015ApJ...802L..25T} introduced a reference frame for the Cen\,A subgroup as the system where two planes of satellites almost lie in the xy-plane, with the normal of the planes corresponding to the z-axis (see also \citealp{2016A&A...595A.119M}). To find a reference system for the M\,101 group we fitted a plane with the help of a singular value decomposition \citep[svd; ][]{1965SJNA....2..205G} at all galaxies lying closer than 1.5\,Mpc from M\,101. {The svd method is a technique generally used in linear algebra. It is an eigendecomposition, where the data will be represented by eigenvectors and eigenvalues, corresponding to a least-square fit to the data.} The resulting sample of eleven galaxies is listed in Table\,\ref{table:fit}. The normal vector of the best fit is given in supergalactic coordinates by $\mathbf{n_{ref}}=(0.6285,-0.0228,-0.7775)$. The plane has a $rms$ thickness of {46\,kpc}. To see how much the distance uncertainties contribute to the fit we ran Monte Carlo simulations where we draw the distance of the galaxy from a normal distribution with a 5\% distance uncertainty as $\sigma$ and the literature distance itself as $\mu$. In every run we compared the angle between the normal of our best fit $\mathbf{n_{ref}}$ with the normal of the run. To determine the contribution of the individual galaxies a second test was conducted. In every run eleven galaxies were randomly drawn from the sample from Table\,\ref{table:fit}, with putting them back in the sample such that some galaxies would not be chosen while others twice etc.\,(Bootstrap test with reshuffle). 
The angle difference in both tests has a maximum of 1.5 degrees which clearly indicates that the best-fitting plane is well defined and can be used as statistically robust reference frame.

\begin{table}[H]
\caption{Galaxies within a $r=1.5$\,Mpc sphere around M\,101 used for the plane fitting.}% title of Table
\label{table:fit}      % is used to refer this table in the text
\centering                          % used for centering table
\begin{tabular}{l c c l c}        % centered columns (4 columns)
\hline\hline                 % inserts double horizontal lines
 & $\alpha_{2000}$ & $\delta_{2000}$ & $D$ & Ref \\    % table heading 
Galaxy Name & (deg) & (deg) &  (Mpc) \\    % table heading 
\hline      \\[-2mm]                  % inserts single horizontal line
NGC\,5195       &         202.4916&       47.2681&       7.66&(1) \\
Holm\,IV       	&         208.6875&       53.9047&       7.24&(2) \\
UGC\,08882      &         209.3083&       54.1008&       8.32& (3)\\
M\,101       	&         210.8000&       54.3505&       6.95& (2)\\
{M101-DF3}        &  		210.7708&	53.6156&	6.52& (4)\\
{M101-DF1}        & 		210.9375&	53.9444&	6.38& (4)\\
NGC\,5474       &         211.2583&       53.6630&       6.98& (2)\\
NGC\,5477       &         211.3875&       54.4608&       6.76& (2)\\
{M101-DF2}        & 		212.1542&	54.3253&	6.87& (4)\\
NGC\,5585       &         214.9500&       56.7303&       5.70& (5)\\
DDO\,194       	&         218.8500&       57.2567&       5.81& (2)\\
\hline\hline 
\end{tabular}
\tablefoot{{Distances are taken from: (1) \cite{2001ApJ...546..681T}, (2) \cite{2013AJ....145..101K}, (3) \cite{2005A&A...437..823R}, { (4) \cite{2017arXiv170204727D}}, and (5) \cite{1994BSAO...38....5K}.}}
\end{table}

We choose the x-axis such that it corresponds to the projection of the line of sight (LoS) onto the plane.  The angle between the LoS to M\,101 and this new x-axis is only 3.6$^{\circ}$, meaning that this plane is lying almost along the LoS. The flat structure extends over 3\,Mpc, showing that the plane is not an artifact of distance uncertainties. The x-axis together with the normal vector $\mathbf{n_{ref}}$ define the reference frame. In order to center M\,101 at its origin one needs to shift the supergalactic coordinates by
$$
v_{SG,M101} =  
v_{SG} +
\begin{pmatrix}
-2.8547 \\
-5.7457  \\
-2.6721  \\
\end{pmatrix} \text{ [Mpc]}
$$ 
The transformation from the shifted supergalactic coordinates to the reference system is then given by the following rotation matrix 
$$\boldsymbol{R}=
 \begin{bmatrix}
   -0.4498 &  -0.8283 & -0.3393\\
   -0.6362 & 0.5630 & -0.5308\\
   0.6285 &-0.0228 &  -0.7775\\
\end{bmatrix}.
$$ 
The final transformation is
$$v_{M101}=\boldsymbol{R}\cdot v_{SG,M101}$$
The best-fitting (reference) plane is shown and discussed below (Sect.\,6.2, Fig.\,7). 

{We note that the geometrical analysis was initially performed before the distance of three additional Dragonfly galaxies were published \citep{2017arXiv170204727D}. Including these new galaxies (DF1, DF2, DF3) the normal of the best-fitting plane changes only by an angle of 0.4 degrees, showing that the plane is statistically robust. With the additional three galaxies the $rms$ thickness of the plane decreased from 49\,kpc to 46\,kpc.}

\section{Analysis and discussion}
In this section we assess the possible membership of the candidates based on their photometric properties, and we analyse the structure of the M\,101 group and the whole complex in the light of the enlarged sample.

\subsection{New dwarf galaxy candidates}
\begin{figure*}[ht]
\centering
  \includegraphics[width=9cm]{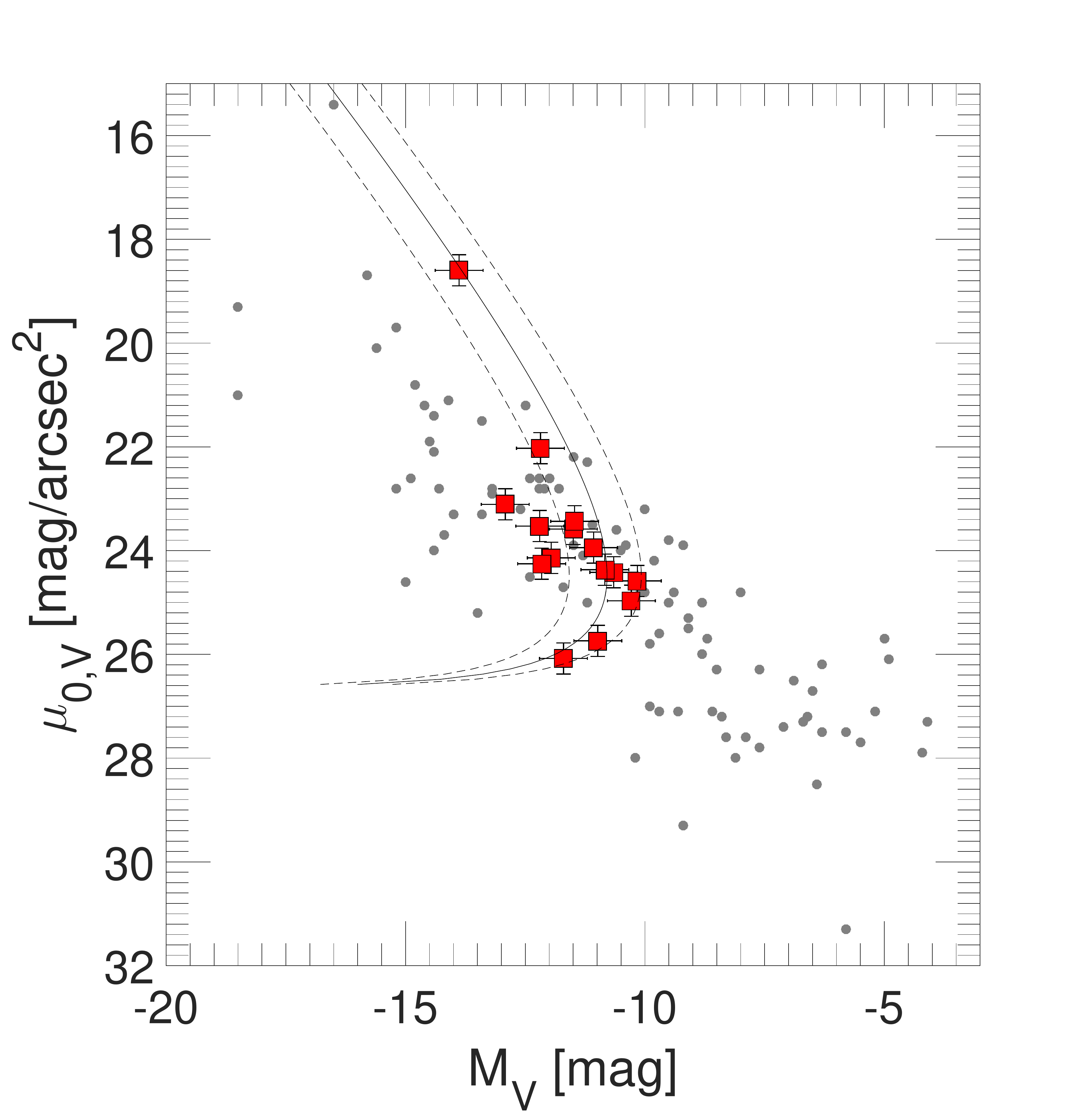}
    \includegraphics[width=9cm]{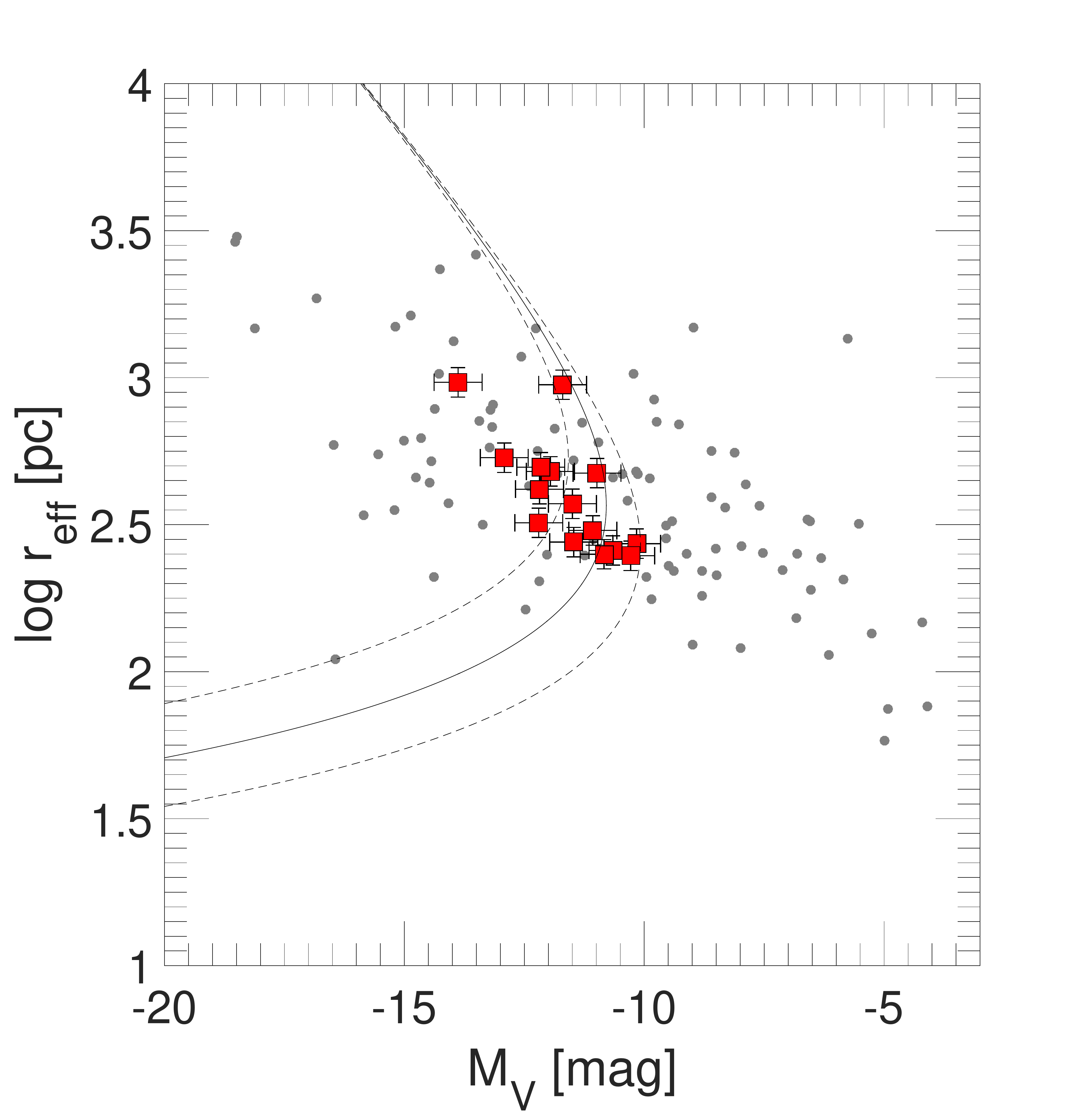}
\caption{Left: $\mu_0$--$M_V$ relation for the photometric parameters of the known Local Group dwarfs \citep[gray dots,][]{2012AJ....144....4M} and the new candidates (red squares). Indicated with the thin line is the assumed completeness boundary (Sect.\,3) at 7\,Mpc, which is bracketed by lines corresponding to assumed distances of 5\,Mpc and 10 Mpc\,, respectively, to take care of the expected distance spread. Right: same data and color coding but this time the $\log r_{eff}$--$M_V$ relation is shown.  }
\label{rel}
\end{figure*}

The usual way to test group membership of dwarf galaxies without direct distance measurements is by comparing their photometric parameters with those of confirmed dwarf galaxies with known distances \citep[e.g.][]{2000AJ....119..593J,2009AJ....137.3009C,2014ApJ...787L..37M,2017A&A...597A...7M}. Dwarf galaxies tend to follow a fairly narrow relation in the central surface brightness--absolute magnitude and the effective radius--absolute magnitude diagram (see Fig.\,\ref{rel}). Note that the central surface brightness is a distance-independent quantity.
If a candidate is a background galaxy not associated with the group or complex the assumed distance for calculating $M_V$ will place the galaxy outside of the relation. In other words, if the parameter values of a candidate fit into the relation with the assumed distance, they are comparable to those of known dwarf galaxies and the candidate can be associated to the group. This convenient test gives us a preliminary, rough handle on the membership status before embarking on a time consuming confirmation by direct distance measurements. 

The performance of the present dwarf candidates in this photometric test is shown in Fig.\,\ref{rel}. Note that we have assumed different distances for our candidates depending on their position relative to the major galaxy they are assigned to (Table 1), roughly 7\,Mpc around M\,101 and 9\,Mpc around M\,51/M\,63. Moreover, given that the dwarfs, even as members of the complex, will be distributed in a large halo around the subgroups, we have to allow for, or expect, a total distance spread of the candidates from $\sim$ 5 to 10 Mpc, giving rise to an additional spread in the photometric relations. The distance spread is also taken care of by overlaying in the relations shown in Fig.\,6 a set of completeness boundaries (cf.\,Sect.\,3) for 5, 7, and 10\,Mpc distance. 

Fig.\,\ref{rel} shows that all but   
one of the new dwarf candidates fit into both of the relations, thus suggesting, or at least being in accord with, their membership in the M\,101 group complex. We note that the outlier is a BCD which does not have to fit into the relations.   

The membership status of dw1343+58 which we classified as Blue Compact Dwarf (BCD) has to be assessed in a different way. Morphologically, the galaxy consists of a high-surface brightness (HSB) irregular central region and an elliptical low-surface brightness component around it, which is 
characteristic for BCDs \citep[e.g.][]{1988A&A...204...10K}. \citet{1996A&A...314...59P,1996A&AS..120..207P} studied the optical structure of BCDs by decomposing their surface brightness profiles into three parts: (i) an underlying extended low-surface brightness component, (ii) an exponential plateau which is mostly seen in iE BCDs, and (iii) an inner HSB region exhibiting a luminosity excess over the plateau which can be fitted with a Gaussian profile. \citet{1996A&A...314...59P}, in their Table\,5, give the 25 mag isophote radius in the $R$-band of the HSB Gaussian component ($P_{25}$) for a sample of BCDs. If we exclude the outliers Haro\,2, Mkn\,297, and I\,Zw\,89, we end up with a mean size of $P_{25}=0.55$\,kpc for BCDs. Now, the estimated $P_{25}$ size of our candidate dw1343+58 is $\sim 15\arcsec$. Assuming $P_{25}=0.55$\,kpc would then put it at a distance of 7.8\,Mpc, which is indeed in accord with M\,101 group membership. We note that this candidate is listed as galaxy in the HYPERLEDA catalog \citep{2003A&A...412...45P}, but not as BCD nor M\,101 dwarf.

As mentioned, the Canes Venatici (CVn) cloud of galaxies is partially overlapping with the M\,101 group complex in sky projection. It is conventionally split at a line-of-sight velocity division line of 400\,km\,s$^{-1}$ ($\sim 5.7$\,Mpc) into the CVn\,I and CVn\,II clouds. CVn\,I cloud members peak at $\sim$300\,km\,s$^{-1}$ ($\sim 4.2$\,Mpc) and CVn\,II cloud members at $\sim$560\,km\,s$^{-1}$ ($\sim 8.0$\,Mpc)\citep{2013AstBu..68..125M}. The whole CVn complex is an extended structure consisting mostly of late-type galaxies of low luminosity and is part of the Coma-Sculptor Cloud, a huge, $\sim$10\,Mpc long, prolate filament, which also includes the Sculptor Cloud, the Local Group, the M\,81 group, and the Cen\,A group \citep{1988ngc..book.....T,2003A&A...398..467K}. In our search area, 11 known galaxies have distances smaller than 5.0\,Mpc, identifying them as part of the CVn I cloud. So, it is conceivable that some of our candidates could in fact be foreground dwarf galaxies. In particular, spiral galaxy M\,94 at a distance of 4.5\,Mpc \citep{2003A&A...398..467K}, one of the major members of CVn I \citep{2013AstBu..68..125M}, is less than 0.5 degrees off our search border (at 12h50m53.5s +41d07m10s). The dwarf candidate dw1255+40 is at a projected distance of only 0.95\,degrees from M\,94, corresponding to a separation of 75\,kpc at the distance of M\,94. Placing this candidate in the vicinity of M\,94 (at 4.5\,Mpc) instead in the M\,101 group complex (at 7 or 9\,Mpc), this would still be fine for the photometric test, i.e.\, the adjusted structural parameters of the candidate would still fit into the relations. Here, a direct distance measurement is needed to confirm its membership in either structure.

\subsection{Structure of the M101 group complex}
\begin{figure*}[ht]
\centering
  \includegraphics[width=18cm]{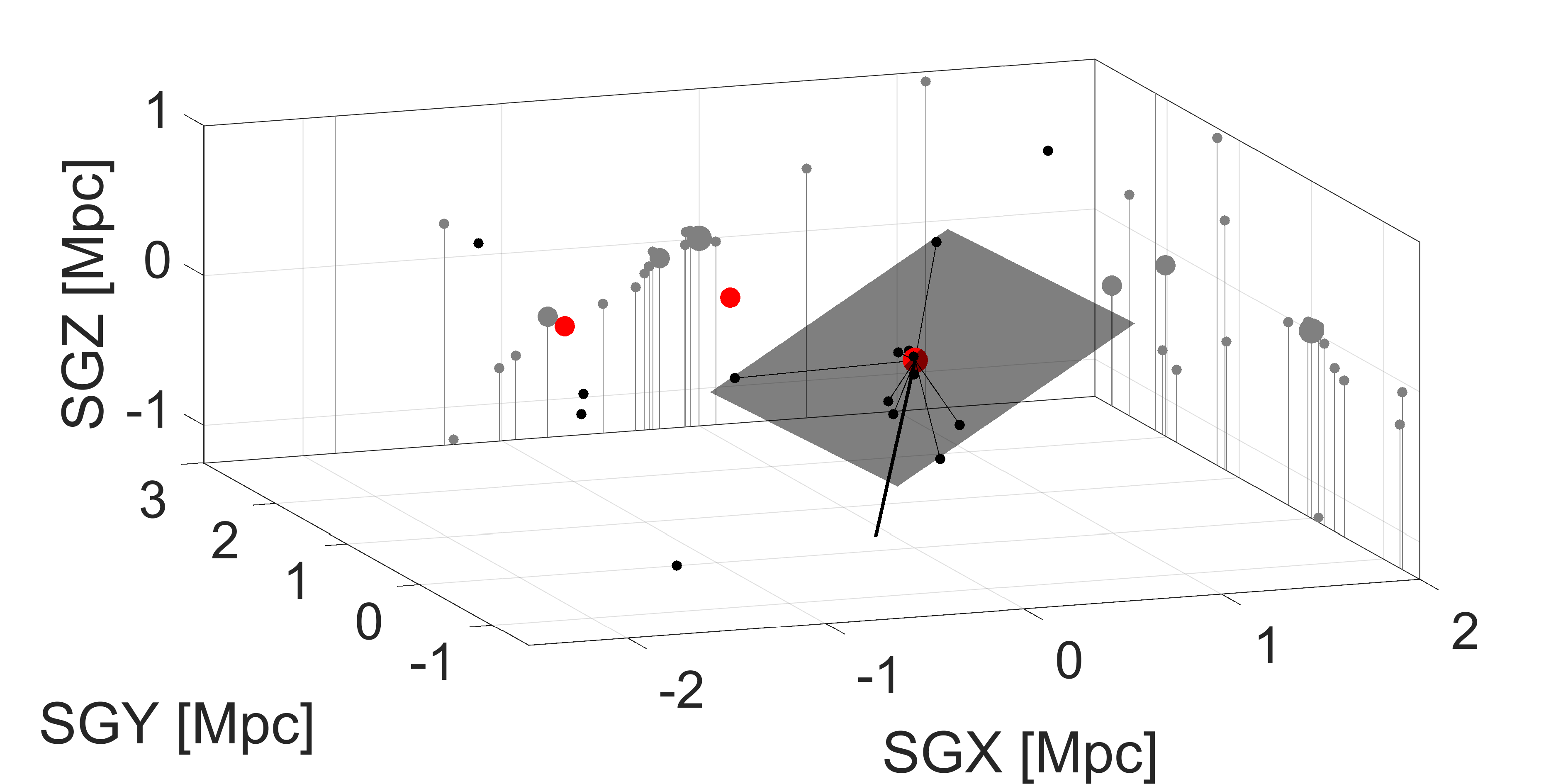}
\caption{3D distribution, in supergalactic coordinates, of all galaxies with distance measurements in the surveyed M\,101 group complex, centered at M\,101. The red dots correspond to the major galaxies M\,101, M\,51 and M\,63, the black dots to dwarf galaxies. The grey dots (shadows) appearing on the SGXSGZ and SGYSGZ--walls are orthogonal projections. The best-fitting plane through the M\,101 subgroup is shown as the grey plane and has a $rms$ of only 46\,kpc. The line of sight between Milky Way and M\,101 is indicated with the thick black line pointing downwards.}
\label{3d}
\end{figure*}

How do these new dwarf galaxy candidates fit into the group complex? We now focus on the structure and geometry of the complex and discuss its impact on the formation history. 
The 15 galaxies with known distances in the survey region are plotted in supergalactic coordinates and centered at M101 in Fig.\,\ref{3d}. Also shown is the best-fitting plane through eight members of the M\,101 subgroup, as calculated in Sect.\,5. The galaxies are cast orthogonally onto the SGXSGZ and SGYSGZ--planes, where they appear as shadows. The highly flattened, filamentary  distribution of the galaxies, especially in the SGXSGZ--plane, is quite striking. The M\,101 plane is a good representation of the whole complex, i.e., the planar structure of the M\,101 subgroup is embedded in a larger flattened structure that encompasses what we call the M\,101 group complex.

\begin{figure*}[ht]
\centering
    \includegraphics[width=9cm]{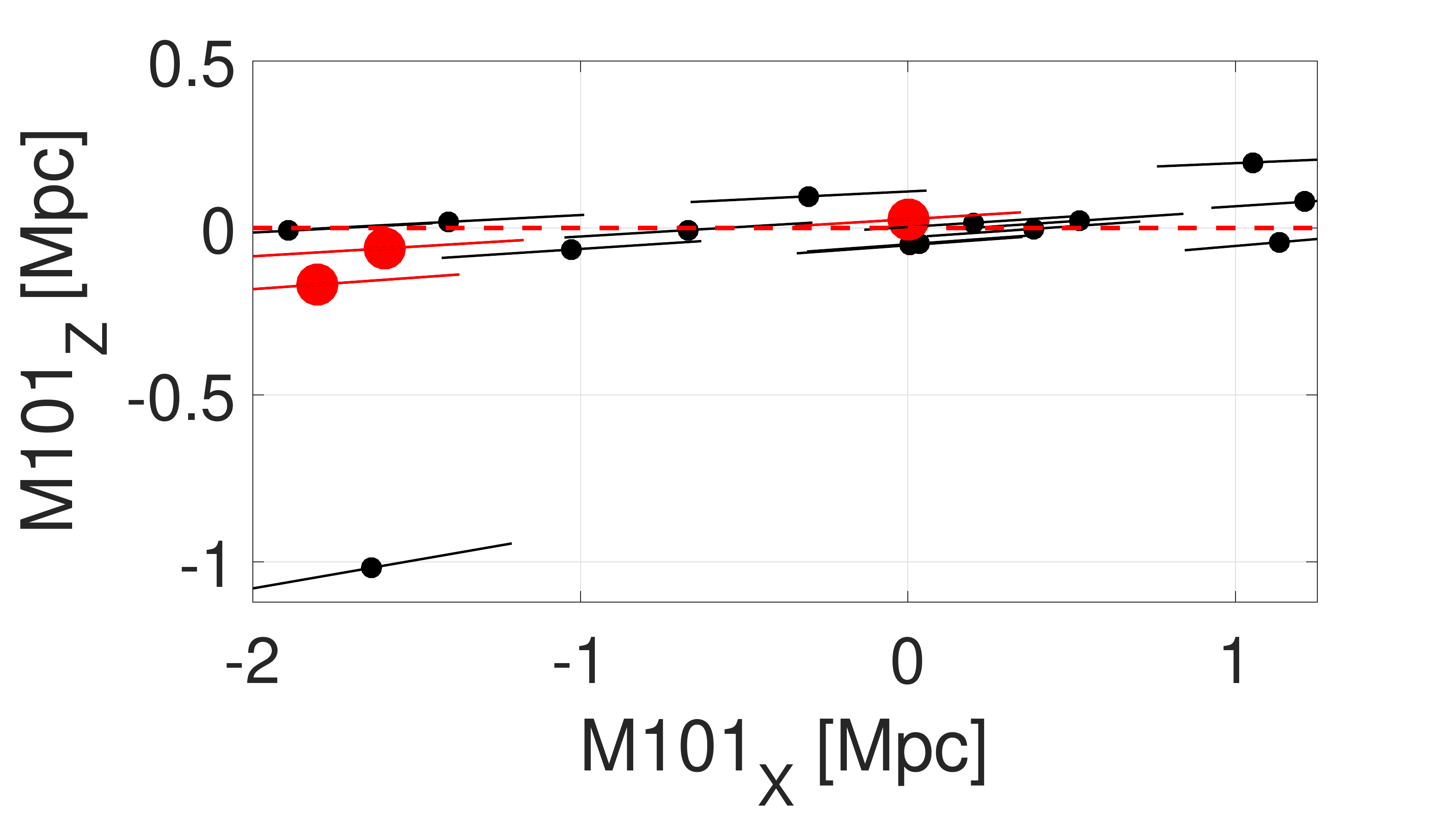}
  \includegraphics[width=9cm]{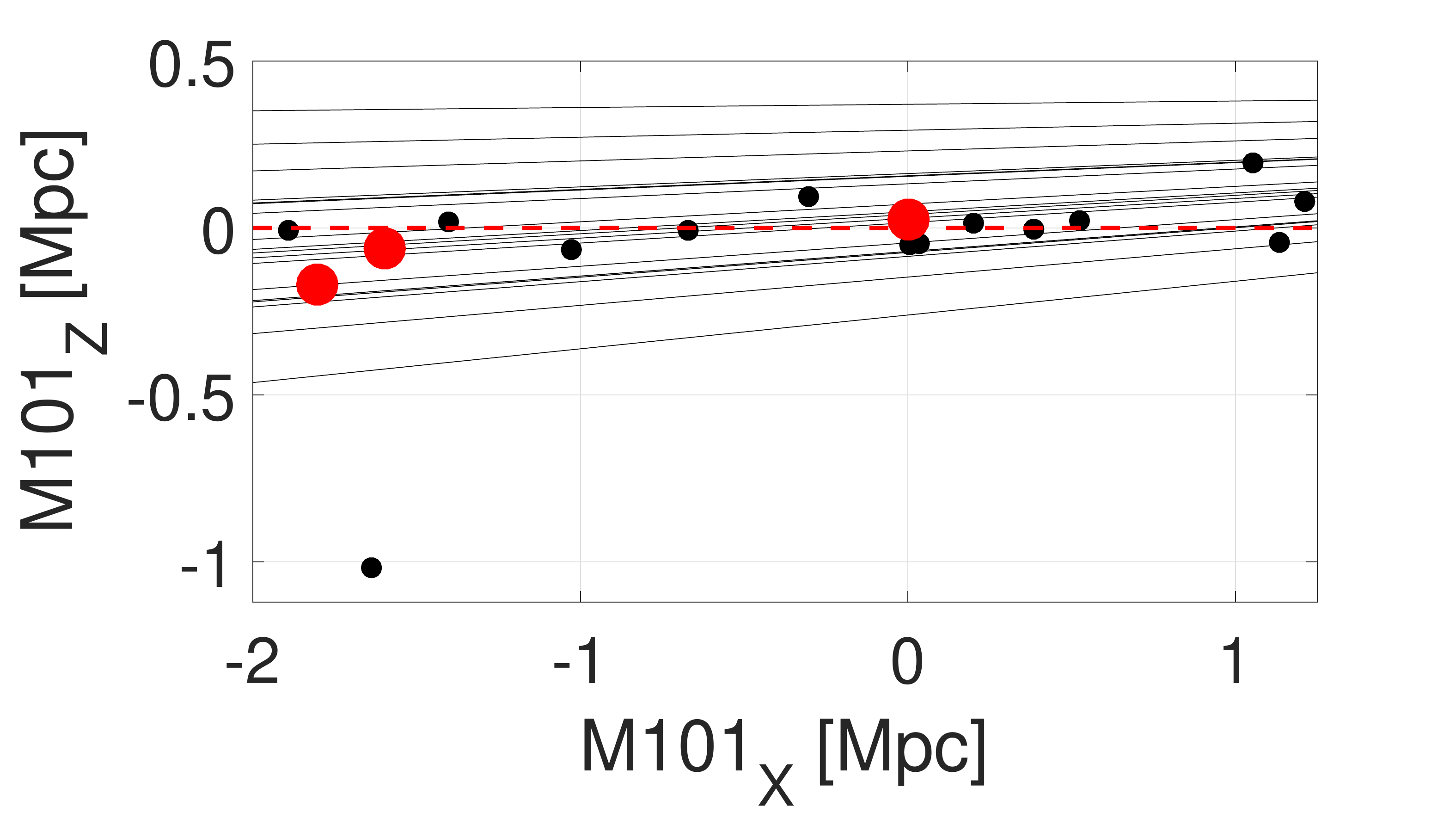} \\
      \includegraphics[width=9cm]{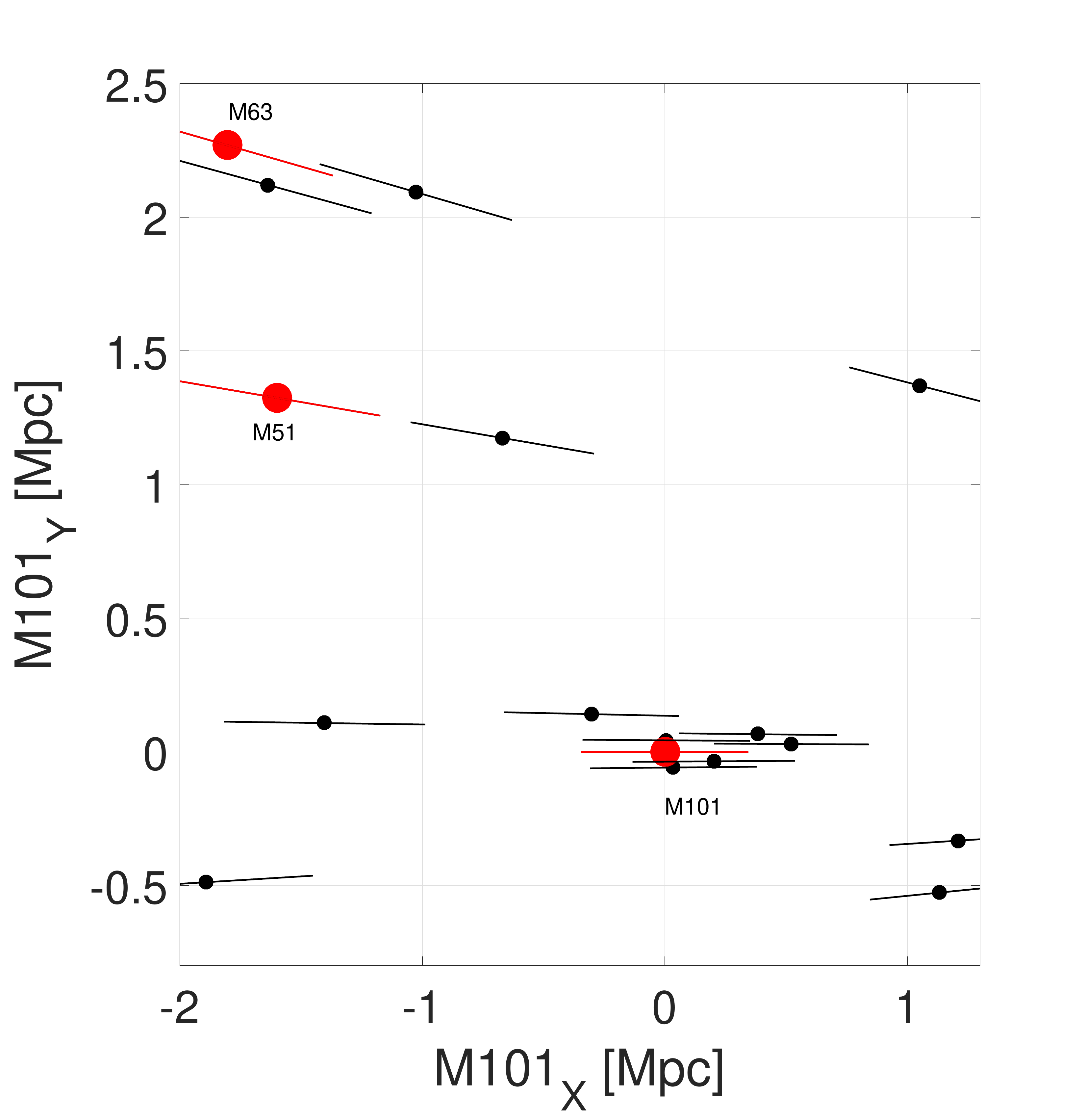}
  \includegraphics[width=9cm]{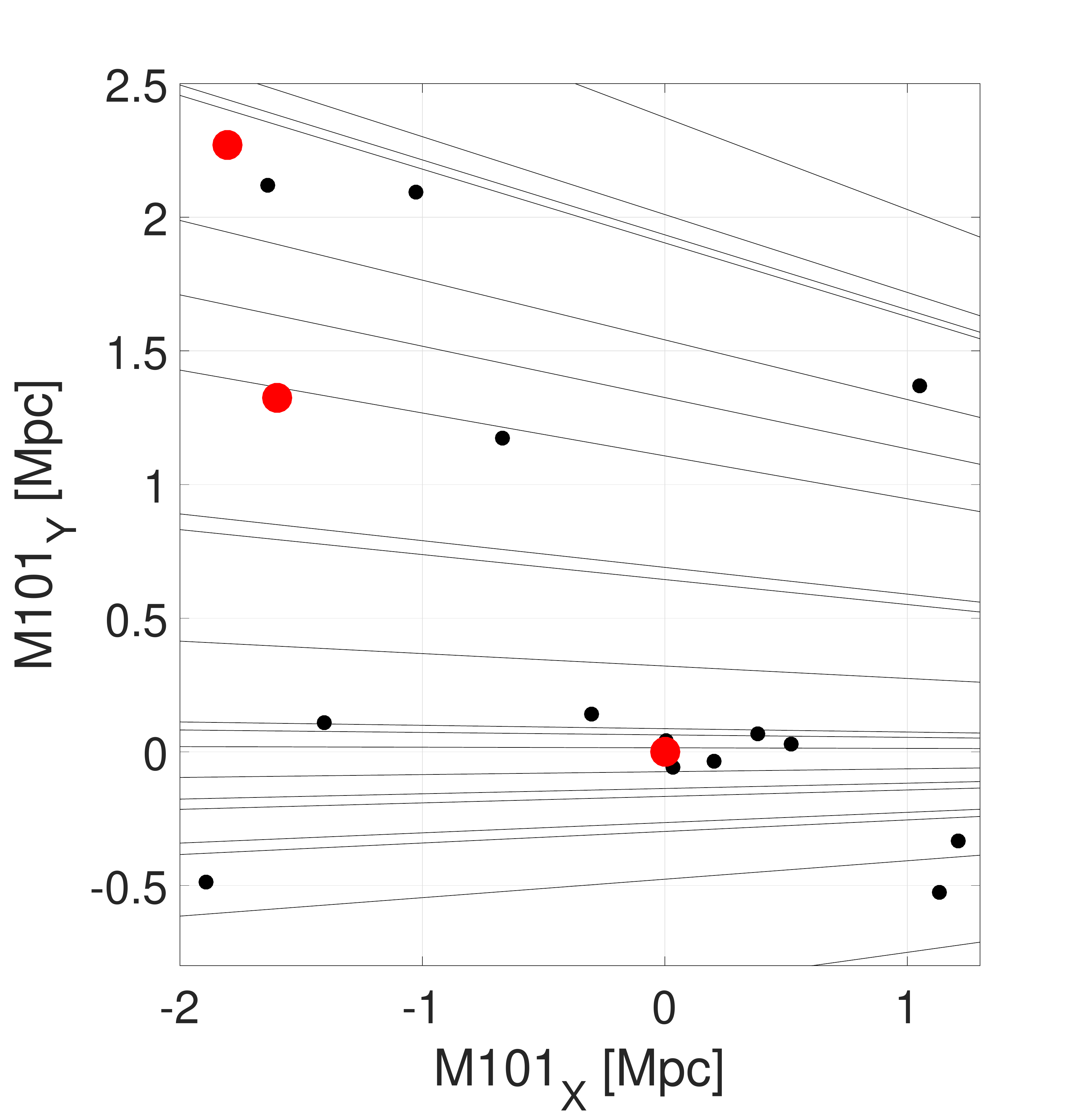} \\
\caption{M101$_{\rm X}$M101$_{\rm Z}$--projection (top) and M101$_{\rm X}$M101$_{\rm Y}$-projection (bottom) of known members and candidate members of the M\,101 group complex in the M\,101 reference system. The top panels essentially give an edge-on view of the complex, the bottom panels a face-on view. Red dots correspond to the three major galaxies M\,101, M\,51, and M\,63, black dots to dwarf galaxies with distances. The thick red dotted line is the best-fitting (reference) plane lying in the M101$_{\rm X}$M101$_{\rm Y}$--plane. The left panels show the known members with 5 percent distance errors indicated by the short lines along the LoS. The right panels additionally give the possible positions of our new dwarf candidates and the candidates from the LV Catalog, in lack of distance measurements shown as long thin black lines again running along their LoS.  }
\label{xz}
\end{figure*}

To further study this flat structure and locate our dwarf candidates in it we now switch to the M\,101 reference frame introduced in Sect.\,5. In this system the best fit corresponds to the M101$_{\rm X}$M101$_{\rm Y}$--plane, which has its origin at (0,0,0). As previously mentioned, {the normal of the best-fitting plane is almost perpendicular to the LoS}. {When this normal is perpendicular to the LoS, then the {LoS} for the dwarf galaxy candidates will be (almost) parallel to the best-fitting plane. As the plane is not perfectly parallel to our view the LoS of the candidates will be systematically shifted along the negative direction of the M101$_{\rm Z}$--axis.} The M101$_{\rm X}$M101$_{\rm Z}$ and M101$_{\rm X}$M101$_{\rm Y}$--projections in this reference system are shown in Fig\,\ref{xz}. In the top left panel the galaxies with known distances and their 5 percent uncertainties are shown. In the top right panel the possible positions of the candidates (dwarfs presented here and the candidates taken from the LV Catalog) are indicated by lines. All lines have a relative shallow slope and cover between 0.06\,Mpc and 0.35\,Mpc in M101$_{\rm Z}$-direction over an interval of 3.25\,Mpc along the M101$_{\rm X}$--axis {(or LoS depth of the M\,101 group complex)}. This narrow spread in  M101$_{\rm Z}$ enables us to study the possible distribution of the candidates without exact knowledge of their distances. All we need are the sky positions and the fact that the M\,101 group complex is flattened almost along the LoS. In the edge-on view one can easily determine whether or not a candidate is part of the filamentary structure. The bottom panels show the structure in the M101$_{\rm X}$M101$_{\rm Y}$--plane, giving a face-on view onto the best-fitting plane. 

Looking at the M101$_{\rm X}$M101$_{\rm Z}$--projection (the edge-on view, top right of Fig\,\ref{xz}) we first verify that almost all known galaxies in the region, notably M\,51 and M\,63, are close to the best-fitting plane through the M\,101 subgroup (the thick red dotted line). That plane through eight members of the subgroup had a $rms$ thickness of {46\,kpc}. If instead a fit is performed at all {16} galaxies lying along the planar structure {(KH\,87, M\,63, M\,51, NGC\,5195, UGC\,08882, KK\,191, Holm\,IV, M\,101, NGC\,5474, DF2, NGC\,5477, DF3, DF1, NGC\,5023, NGC\,5585, and DDO\,194), we calculate a $rms=67$\,kpc}, which is still remarkably thin. Only DDO\,182 falls outside of the structure.
Moreover, it is clearly visible that most of the candidate dwarfs lie within (or near) the flattened structure outlined by the know members of the M\,101 group complex. 

Looking at the M101$_{\rm X}$M101$_{\rm Y}$--projection (face-on view, bottom right) we note that only three candidates lie in the space between M\,101 and M\,51 (ignore the lines which are close to M\,101). This is further evidence that the M\,101 and M\,51/M\,63 subgroups form separated groups as suggested by \cite{2015AstL...41..239T}. Most new dwarf candidates are in the direction of M\,101.  

With a length of over 3\,Mpc this flattened structure could be attributed to the Cosmic Web where the galaxies are known to be aligned along dark matter filaments. The question, then, is how this structure compares to other filamentary structures. \citet{2010MNRAS.409..156B} identified individual filamentary structures in SDSS and compared their properties to those produced in cosmological simulations. They found a mean SDSS filament width of $5.5\pm 1.1$ or $8.4\pm 1.4$ $h^{-1}$\,Mpc, depending on the smoothing length, which in size is comparable to our best-fitting plane ($\sim$ 3 Mpc) when the bottom panel of Fig.\,\ref{xz} is taken as a measure. However, the thickness of the M\,101 complex, of only $rms$=67\,kpc is remarkable. This is a factor of $\sim$ 40 smaller than the size of the structure. 

A direct comparison with simulations is difficult. \citet{2010MNRAS.407.1449G} used an algorithm to identify and analyze filaments in cosmological simulations. They plotted the filament thickness as a function of the filament length with a bin size of 10\,Mpc. In the first bin (0 to 10\,Mpc) a median thickness of 1.3\,Mpc is estimated. The problem with this result for the purpose of comparison is the low resolution of filament length steps. In their Fig. 8 (upper right panel) they present the count of filaments as a function of thickness. There is a small signal at 0.1\,Mpc h$^{-1}$ but it remains unclear if this is due to the interpolation between zero and the first point of the function, rather than being a real signal. Even when we assume that it is a real signal, the probability of a thickness as small as observed in the M\,101 group complex is essentially zero. If the M\,101 group complex is the usual type of a large-scale filamentary structure, its small thickness has to be explained.

One possible explanation for this special configuration is given by the presence of the nearby Local Void (see Courtois et al.\,2013). The planar M\,101 group complex is well-aligned at the edge of the Local Void and can be seen as part of its boundary. The formation of the flattened structure itself could be induced by the expansion of the Local Void.

Regarding the thin planar structure of the M\,101 subgroup itself, with its very small $rms$ thickness of 46\,kpc, the question arises whether this could be a similar phenomenon as the local planes: the Vast Polar Structure (VPOS) of the Milky Way \citep{2012MNRAS.423.1109P} or the Great Plane of Andromeda \citep[GPoA,][]{2013Natur.493...62I}. Remember that much of the motivation to find new faint dwarf galaxies outside the LG is precisely to look for analoguous structures, because the local planes are a challenge to the standard $\Lambda$CDM scenario of structure formation (e.g., \citealp{2014MNRAS.442.2362P,2016arXiv161201529C}). However, we believe that the M\,101 plane is not the same phenomenon, even though the M\,101 plane is only a factor of two or three thicker than the local planes. The scales and the objects that define the planes are different. The local planes are defined by very faint and ultra-faint dwarf satellites in the immediate vicinity of their host galaxies (closer than the virial radius of 250\,kpc). In contrast, the M\,101 plane is defined by still fairly luminous dwarf galaxies at separations as large as 1\,Mpc. Still, the flattened structure of the M\,101 subgroup -- and its extension over the whole complex -- is remarkable.  

How is the galaxy 
M\,101 itself inclined to the group plane?
As mentioned in Sect.\,4 the spiral galaxy M\,101 is seen face-on, which means that our LoS essentially coincides with the direction of the disk normal $\mathbf{n_{M101}}=(0.4107,0.8267,0.3845)$ in supergalactic cartesian coordinates. Hence, with an angle of only 3.6$^{\circ}$ between $\mathbf{n_{M101}}$ and the planar structure of the group, the disk of M\,101 is perpendicular and its normal parallel to the plane. \citet{2014MNRAS.444.1453D} studied the alignment of galaxy spins within the Cosmic Web with a large-scale hydrodynamical cosmological simulation and found that more massive galaxies tend to have their spin direction perpendicular to a filament, while less massive galaxies have their spin direction preferentially parallel to a filament, with a transition mass around $M_{*} \approx 3 \times 10^{10}$ M$_{\odot}$. \citet{2015AstL...41..239T} calculated a dynamical mass for M\,101 of $6.2 \times 10^{11}$\,M$_{\odot}$ and gave a M/L ratio of 18, thus the stellar mass of M\,101 would be $3.4 \times 10^{10}$\,M$_{\odot}$, assuming M/L$=1$ for the stellar component. Hence, the stellar mass of M\,101 is just around the Dubois transition mass and either way is in accord with this work. A follow-up study based on the same simulation framework \citep{2014MNRAS.445L..46W} revealed a strong correlation between the merger history and the spin alignment: the more mergers contribute to the mass of a galaxy, the more likely its spin will be perpendicular to the filament. In contrast, the spin of galaxies with no merger is more likely aligned with the filament. This would then suggest that M\,101 has undergone little or no mergers in its formation history. A view which is well in accord with the observation that M\,101 has a small (or essentially absent) bulge, for in the standard model of bottom-up structure formation bulges are formed in merger events \citep[e.g.][and references therein]{2016ASSL..418..317B}. 

The lack of a strong bulge in M\,101 is an important observation in itself.
\citet{2010ApJ...723...54K} pointed out the challenge of bulgeless spiral galaxies for hierarchical formation scenarios. How can such massive spiral galaxies like M\,101 form out of merger events without growing a prominent bulge? \citet{2016ApJ...817...75L} showed that there is a correlation between bulge size and the number of tidal dwarf galaxies (N$_S$). However, in a $\Lambda$CDM scenario there should be no correlation between these two quantities, because N$_S$ is driven by the dark matter mass of the host galaxy and not by its formation history \citep{2010A&A...523A..32K}. In a generalized model of gravity without DM one does expect such a correlation between the bulge and N$_S$, because (tidal dwarf) satellites form in rare fly-by encounters. Bulges themselves would also form in such encounters, making the bulge-to-disk ratio a measure for past interactions \citep{2016ApJ...817...75L}. What would such a scenario predict for the M\,101 group? As M\,101 is a spiral galaxy without a bulge only few or even no dwarf spheroidals should exist in the group. \cite{2005nfcd.conf..295K} reported that bulgeless galaxies generally have no or only few known dwarf spheroidal companions. Up to this day only  {three early-type dwarfs} have been confirmed as members of the M\,101 group. While in a generalized gravity scenario this missing dSph problem is well explained, the standard model of cosmology needs to find a mechanism for the low abundance of dwarf spheroidals around bulgeless spirals. 

How do our new dwarf detections fit into this picture? This can be evaluated by way of a comparison with the Andromeda subgroup. In \citet{2009AJ....137.3009C} the authors show the cumulative luminosity functions (LF) of the Cen\,A, M\,81 and Andromeda satellites. Our survey reached a limiting magnitude of $M_V\sim -10$\,mag, assuming a distance of 7\,Mpc. Among the three satellite populations the cumulative LF of the Andromeda subgroup shows the lowest abundance, with 15 satellites down to $M_V= -10$\,mag. Andromeda and M\,101 also have a similar total $B$-band luminosity ($\Delta B\approx 0.5$\,mag), thus a comparison is reasonable. Within a projected virial radius of $\approx 260$\,kpc around M\,101 \citep{2014ApJ...787L..37M}, comparable to {Andromeda} with 230\,kpc, there are {seven} confirmed members (Holm\,IV, UGC\,08882, DF3, DF1, NGC\,5474, NGC\,5477, DF2) down to $V$ magnitude of $-10$ {(three early-type and four late-type dwarfs). Additionally there are five dwarf candidates within this radius (M101\,dwD, M101\,dwC, M101\,dwA, M101\,dwB, dw1412+56). Assuming a distance of 6.95\,Mpc for all of them, five additional members would contribute to the population, giving a total of twelve, putting the M\,101 subgroup almost on a par with the Andromeda subgroup. On the other hand, assuming a positive detection rate of 60\,\% (14 out of 22 candidates of the M\,81 group were confirmed as members \citep{2013AJ....146..126C}, the rest being background or cirrus) we would gain only three additional members, adding to a total of ten satellites. This, in turn, would indeed indicate a smaller population of dwarf galaxies in the M\,101 subgroup. Using all distance data available and calculating the 3D distances to M\,101, only three (NGC\,5474, NGC\,5477, DF2) out of the seven galaxies lying in the projected virial radius are closer than 260\,kpc to M\,101, drastically increasing the missing-satellite problem in the M\,101 subgroup.  Similarly, \citet{2017arXiv170204727D} draw the same conclusion with their recent publication of HST data for their Dragonfly candidates. The authors also cautiously predict a too-big-to-fail problem for the M\,101 subgroup, based on the low abundance of bright dwarf satellites around M\,101.}
Clearly, we need distance measurements for the dwarf candidates in the vicinity of M\,101 to answer the question whether, as claimed, the abundance of dwarf satellites of M\,101 is exceptionally low, {hence a missing-satellite problem really exists in the M\,101 subgroup. In this context it is noteworthy that there are almost no new candidates in the virial radii of all three host galaxies. Could we face similar problems in the M\,51 and M\,63 subgroups?} 

The alignment of the spin vector with the planar structure and the low number of M\,101 dwarf satellites -- if confirmed -- lead to the conclusion that M\,101 has a weak merger history. {Additional evidence for this is given by \citet{2014ApJ...782L..24V} who point out a lack of a stellar halo of M\,101. Such stellar halos are formed from debris of shredded satellite galaxies \citep{2009Natur.461...66M}, hence are an indicator for previous interaction.} Do all these evidences mean that there is no evidence for interaction in the galaxy group? \citet{2013ApJ...762...82M} studied the faint outskirts of the spiral up to a limiting surface brightness (star density) of $\mu_B\sim 29.5$\,mag arcsec$^{-2}$ and found no evidence of extended stellar tidal tails around M\,101 or its companions. Such tails should be expected when M\,101 had a recent encounter with one of its massive companions. However, two low-surface brightness features were found in the outer disk. One of them must have formed very recently due to its blue stellar population. The authors argue that this faint blue feature could have formed in fly-by encounters with NGC\,5477 and NGC\,5474. The latter galaxy exhibits an off-centered central bulge, suggesting some interaction in the past. The high velocity gas in the disk of M\,101 is another indicator for tidal interaction, possibly with the companion NGC\,5477 \citep{1991A&A...243..109C}.

{More prominent than M\,101 in terms of interaction is the ongoing merger between M\,51 and NGC\,5195 \citep{1972BAAS....4..214T}. In \citet{2010ASPC..423..240D} this merger was simulated with a hydrodynamical model with a highly elliptical orbit, where NGC\,5195 passes trough the disc of M\,51 twice. A qualitative assessment of the trajectory of NGC\,5195 shows that it correlates with our best-fitting plane, which is not surprising, as accretion happens along the filament \citep{2014MNRAS.443.1274L}.}

In contrast to more distant dwarf galaxy candidates \citep[e.g. ][d\,>\,20\,Mpc]{2016MNRAS.463.1284O,2016A&A...596A..23S}, the new dwarf galaxy candidates in the M\,101 group complex can be resolved into stars with appropriate equipment from the ground (e.g. Subaru) or in space (HST). The task is to confirm these objects as nearby stellar systems, excluding the possibility of being more distant, unresolved galaxies or Galactic cirrus, Measuring their distances also allows to allocate each of them to one of the three subgroups in the M\,101 group complex (or the Canes Venatici I cloud in the foreground). Will the candidates spread along the 3\,Mpc sheet or are they clustered around the main galaxies M\,101, M\,51, and M\,63? Accurate distance measurements will be key for the study of the fine structure of large-scale structure in the M\,101 group filament. 

\section{Conclusion}
In this work we presented the results of a dwarf galaxy search covering the M\,101 group of galaxies and its wider environment including M\,51 and M\,63 with publicly available data from the Sloan Digital Sky Survey. We searched a sky area of 330 square degrees and found 15 new dwarf candidates. Surface photometry was performed for all candidates in the $gr$ bands and S\'ersic profiles were fitted to the surface brightness profiles. We tested the group membership with the classical tools at hand: the central surface brightness--absolute magnitude and effective radius--absolute magnitude relations. The candidates indeed fit in comparison to the structural parameters of known Local Group dwarf galaxies, making them good candidates of the M\,101 group complex. Distance measurements are nevertheless needed to confirm these results. We discussed the possibility that some of the candidates could be dwarf members of the Canes Venatici cloud in the near foreground. 

The second part of this work was committed to the 3D spatial distribution of the group and the whole complex. We found that all but one of the galaxies with known distances lie in a thin plane with {$rms=67$\,kpc} and a length of over 3\,Mpc, including M\,51 and M\,63. The plane was defined by a best fit at the M\,101 subgroup alone, i.e., M\,101 and its neighbours within 1.5\,Mpc, with a $rms$ thickness of only  {46\,kpc}. {The recent publication of three additional dwarf galaxies \citep{2017arXiv170204727D} strengthens the picture of a thin, planar structure. This structure} happens to be well aligned with our line of sight, giving us the opportunity to place the new dwarf candidates relative to this plane without knowing their exact distances. For this we defined a M\,101 reference frame where the z-axis corresponds to the normal of the plane. 

The flattened structure of the M\,101 group complex is aligned with the envelope of the Tully Void which could explain its formation by the expansion of the void. There is a clear alignment between the spin direction of M\,101 and the planar structure: the spiral disk of M\,101 is almost perpendicular to the best-fitting plane. In a $\Lambda$CDM scenario this can be explained by a weak merger history. The missing bulge of M\,101 also strengthens the case for a merger-less formation history. We discussed the impact of such a formation history with the abundance of dwarf spheroidals in a $\Lambda$CDM and a generalized gravity scenario.

Future distance measurements of the candidates in the M\,101 group complex will give us answers to the questions whether the planar structure is only a small number statistics artifact or a real cosmic structure, and if the latter is true, how thin it is and what are the implications? Will the candidates cluster around the main galaxies or are they more widely distributed along this filament?

\begin{acknowledgements}
OM and BB are grateful to the Swiss National Science Foundation for financial support. HJ acknowledges the support of the Australian Research Council through Discovery Project DP150100862. The authors thank Marcel Pawlowski and Marina Rejkuba for interesting discussions and helpful input.
{The authors like to thank the anonymous referee for helpful comments that improved the paper.}
\end{acknowledgements}

\bibliographystyle{aa}
\bibliography{aanda}

\begin{thebibliography}{77}
\expandafter\ifx\csname natexlab\endcsname\relax\def\natexlab#1{#1}\fi

\bibitem[{{Alam} {et~al.}(2015){Alam}, {Albareti}, {Allende Prieto}, {Anders},
  {Anderson}, {Anderton}, {Andrews}, {Armengaud}, {Aubourg}, {Bailey}, \&
  et~al.}]{2015ApJS..219...12A}
{Alam}, S., {Albareti}, F.~D., {Allende Prieto}, C., {et~al.} 2015, \apjs, 219,
  12

\bibitem[{{Belokurov} {et~al.}(2010){Belokurov}, {Walker}, {Evans}, {Gilmore},
  {Irwin}, {Just}, {Koposov}, {Mateo}, {Olszewski}, {Watkins}, \&
  {Wyrzykowski}}]{2010ApJ...712L.103B}
{Belokurov}, V., {Walker}, M.~G., {Evans}, N.~W., {et~al.} 2010, \apjl, 712,
  L103

\bibitem[{{Binggeli}(1989)}]{1989ASSL..151...47B}
{Binggeli}, B. 1989, in Astrophysics and Space Science Library, Vol. 151, Large
  Scale Structure and Motions in the Universe, ed. M.~{Mezzetti},
  G.~{Giuricin}, F.~{Mardirossian}, \& M.~{Ramella}, 47--61

\bibitem[{{Binggeli} {et~al.}(1987){Binggeli}, {Tammann}, \&
  {Sandage}}]{1987AJ.....94..251B}
{Binggeli}, B., {Tammann}, G.~A., \& {Sandage}, A. 1987, \aj, 94, 251

\bibitem[{{Bond} {et~al.}(2010){Bond}, {Strauss}, \&
  {Cen}}]{2010MNRAS.409..156B}
{Bond}, N.~A., {Strauss}, M.~A., \& {Cen}, R. 2010, \mnras, 409, 156

\bibitem[{{Bremnes} {et~al.}(1999){Bremnes}, {Binggeli}, \&
  {Prugniel}}]{1999A&AS..137..337B}
{Bremnes}, T., {Binggeli}, B., \& {Prugniel}, P. 1999, \aaps, 137, 337

\bibitem[{{Brooks} \& {Christensen}(2016)}]{2016ASSL..418..317B}
{Brooks}, A. \& {Christensen}, C. 2016, Galactic Bulges, 418, 317

\bibitem[{{Carrillo} {et~al.}(2017){Carrillo}, {Bell}, {Bailin}, {Monachesi},
  {de Jong}, {Harmsen}, \& {Slater}}]{2017MNRAS.465.5026C}
{Carrillo}, A., {Bell}, E.~F., {Bailin}, J., {et~al.} 2017, \mnras, 465, 5026

\bibitem[{{Cautun} {et~al.}(2015){Cautun}, {Bose}, {Frenk}, {Guo}, {Han},
  {Hellwing}, {Sawala}, \& {Wang}}]{2015MNRAS.452.3838C}
{Cautun}, M., {Bose}, S., {Frenk}, C.~S., {et~al.} 2015, \mnras, 452, 3838

\bibitem[{{Cautun} \& {Frenk}(2016)}]{2016arXiv161201529C}
{Cautun}, M. \& {Frenk}, C.~S. 2016, ArXiv e-prints
  [\eprint[arXiv]{1612.01529}]

\bibitem[{{Chiboucas} {et~al.}(2013){Chiboucas}, {Jacobs}, {Tully}, \&
  {Karachentsev}}]{2013AJ....146..126C}
{Chiboucas}, K., {Jacobs}, B.~A., {Tully}, R.~B., \& {Karachentsev}, I.~D.
  2013, \aj, 146, 126

\bibitem[{{Chiboucas} {et~al.}(2009){Chiboucas}, {Karachentsev}, \&
  {Tully}}]{2009AJ....137.3009C}
{Chiboucas}, K., {Karachentsev}, I.~D., \& {Tully}, R.~B. 2009, \aj, 137, 3009

\bibitem[{{Combes}(1991)}]{1991A&A...243..109C}
{Combes}, F. 1991, \aap, 243, 109

\bibitem[{{Courtois} {et~al.}(2013){Courtois}, {Pomar{\`e}de}, {Tully},
  {Hoffman}, \& {Courtois}}]{2013AJ....146...69C}
{Courtois}, H.~M., {Pomar{\`e}de}, D., {Tully}, R.~B., {Hoffman}, Y., \&
  {Courtois}, D. 2013, \aj, 146, 69

\bibitem[{{Crnojevi{\'c}} {et~al.}(2014){Crnojevi{\'c}}, {Sand}, {Caldwell},
  {Guhathakurta}, {McLeod}, {Seth}, {Simon}, {Strader}, \&
  {Toloba}}]{2014ApJ...795L..35C}
{Crnojevi{\'c}}, D., {Sand}, D.~J., {Caldwell}, N., {et~al.} 2014, \apjl, 795,
  L35

\bibitem[{{Crnojevi{\'c}} {et~al.}(2016){Crnojevi{\'c}}, {Sand}, {Spekkens},
  {Caldwell}, {Guhathakurta}, {McLeod}, {Seth}, {Simon}, {Strader}, \&
  {Toloba}}]{2016ApJ...823...19C}
{Crnojevi{\'c}}, D., {Sand}, D.~J., {Spekkens}, K., {et~al.} 2016, \apj, 823,
  19

\bibitem[{{Danieli} {et~al.}(2017){Danieli}, {van Dokkum}, {Merritt},
  {Abraham}, {Zhang}, {Karachentsev}, \& {Makarova}}]{2017arXiv170204727D}
{Danieli}, S., {van Dokkum}, P., {Merritt}, A., {et~al.} 2017, ArXiv e-prints
  [\eprint[arXiv]{1702.04727}]

\bibitem[{{Dobbs} {et~al.}(2010){Dobbs}, {Theis}, {Pringle}, \&
  {Bate}}]{2010ASPC..423..240D}
{Dobbs}, C.~L., {Theis}, C., {Pringle}, J.~E., \& {Bate}, M.~R. 2010, in
  Astronomical Society of the Pacific Conference Series, Vol. 423, Galaxy Wars:
  Stellar Populations and Star Formation in Interacting Galaxies, ed.
  B.~{Smith}, J.~{Higdon}, S.~{Higdon}, \& N.~{Bastian}, 240

\bibitem[{{Dubois} {et~al.}(2014){Dubois}, {Pichon}, {Welker}, {Le Borgne},
  {Devriendt}, {Laigle}, {Codis}, {Pogosyan}, {Arnouts}, {Benabed}, {Bertin},
  {Blaizot}, {Bouchet}, {Cardoso}, {Colombi}, {de Lapparent}, {Desjacques},
  {Gavazzi}, {Kassin}, {Kimm}, {McCracken}, {Milliard}, {Peirani}, {Prunet},
  {Rouberol}, {Silk}, {Slyz}, {Sousbie}, {Teyssier}, {Tresse}, {Treyer},
  {Vibert}, \& {Volonteri}}]{2014MNRAS.444.1453D}
{Dubois}, Y., {Pichon}, C., {Welker}, C., {et~al.} 2014, \mnras, 444, 1453

\bibitem[{{Ferguson}(1990)}]{1990PhDT.........1F}
{Ferguson}, H.~C. 1990, PhD thesis, Johns Hopkins Univ., Baltimore, MD.

\bibitem[{{Ferguson} \& {Sandage}(1988)}]{1988AJ.....96.1520F}
{Ferguson}, H.~C. \& {Sandage}, A. 1988, \aj, 96, 1520

\bibitem[{{Golub} \& {Kahan}(1965)}]{1965SJNA....2..205G}
{Golub}, G. \& {Kahan}, W. 1965, SIAM Journal on Numerical Analysis, 2, 205

\bibitem[{{Gonz{\'a}lez} \& {Padilla}(2010)}]{2010MNRAS.407.1449G}
{Gonz{\'a}lez}, R.~E. \& {Padilla}, N.~D. 2010, \mnras, 407, 1449

\bibitem[{{Gunn} {et~al.}(2006){Gunn}, {Siegmund}, {Mannery}, {Owen}, {Hull},
  {Leger}, {Carey}, {Knapp}, {York}, {Boroski}, {Kent}, {Lupton}, {Rockosi},
  {Evans}, {Waddell}, {Anderson}, {Annis}, {Barentine}, {Bartoszek}, {Bastian},
  {Bracker}, {Brewington}, {Briegel}, {Brinkmann}, {Brown}, {Carr},
  {Czarapata}, {Drennan}, {Dombeck}, {Federwitz}, {Gillespie}, {Gonzales},
  {Hansen}, {Harvanek}, {Hayes}, {Jordan}, {Kinney}, {Klaene}, {Kleinman},
  {Kron}, {Kresinski}, {Lee}, {Limmongkol}, {Lindenmeyer}, {Long}, {Loomis},
  {McGehee}, {Mantsch}, {Neilsen}, {Neswold}, {Newman}, {Nitta}, {Peoples},
  {Pier}, {Prieto}, {Prosapio}, {Rivetta}, {Schneider}, {Snedden}, \&
  {Wang}}]{2006AJ....131.2332G}
{Gunn}, J.~E., {Siegmund}, W.~A., {Mannery}, E.~J., {et~al.} 2006, \aj, 131,
  2332

\bibitem[{{Ibata} {et~al.}(2013){Ibata}, {Lewis}, {Conn}, {Irwin},
  {McConnachie}, {Chapman}, {Collins}, {Fardal}, {Ferguson}, {Ibata}, {Mackey},
  {Martin}, {Navarro}, {Rich}, {Valls-Gabaud}, \&
  {Widrow}}]{2013Natur.493...62I}
{Ibata}, R.~A., {Lewis}, G.~F., {Conn}, A.~R., {et~al.} 2013, NAT, 493, 62

\bibitem[{{Jacobs} {et~al.}(2009){Jacobs}, {Rizzi}, {Tully}, {Shaya},
  {Makarov}, \& {Makarova}}]{2009AJ....138..332J}
{Jacobs}, B.~A., {Rizzi}, L., {Tully}, R.~B., {et~al.} 2009, \aj, 138, 332

\bibitem[{{Javanmardi} {et~al.}(2016){Javanmardi}, {Martinez-Delgado},
  {Kroupa}, {Henkel}, {Crawford}, {Teuwen}, {Gabany}, {Hanson}, {Chonis}, \&
  {Neyer}}]{2016A&A...588A..89J}
{Javanmardi}, B., {Martinez-Delgado}, D., {Kroupa}, P., {et~al.} 2016, \aap,
  588, A89

\bibitem[{{Jerjen} {et~al.}(2000){Jerjen}, {Binggeli}, \&
  {Freeman}}]{2000AJ....119..593J}
{Jerjen}, H., {Binggeli}, B., \& {Freeman}, K.~C. 2000, \aj, 119, 593

\bibitem[{{Karachentsev} {et~al.}(2004){Karachentsev}, {Karachentseva},
  {Huchtmeier}, \& {Makarov}}]{2004AJ....127.2031K}
{Karachentsev}, I.~D., {Karachentseva}, V.~E., {Huchtmeier}, W.~K., \&
  {Makarov}, D.~I. 2004, \aj, 127, 2031

\bibitem[{{Karachentsev} {et~al.}(2005){Karachentsev}, {Karachentseva}, \&
  {Sharina}}]{2005nfcd.conf..295K}
{Karachentsev}, I.~D., {Karachentseva}, V.~E., \& {Sharina}, M.~E. 2005, in IAU
  Colloq. 198: Near-fields cosmology with dwarf elliptical galaxies, ed.
  H.~{Jerjen} \& B.~{Binggeli}, 295--302

\bibitem[{{Karachentsev} {et~al.}(1994){Karachentsev}, {Kopylov}, \&
  {Kopylova}}]{1994BSAO...38....5K}
{Karachentsev}, I.~D., {Kopylov}, A.~I., \& {Kopylova}, F.~G. 1994, Bulletin of
  the Special Astrophysics Observatory, 38, 5

\bibitem[{{Karachentsev} {et~al.}(2013){Karachentsev}, {Makarov}, \&
  {Kaisina}}]{2013AJ....145..101K}
{Karachentsev}, I.~D., {Makarov}, D.~I., \& {Kaisina}, E.~I. 2013, \aj, 145,
  101

\bibitem[{{Karachentsev} {et~al.}(2003){Karachentsev}, {Sharina}, {Dolphin},
  {Grebel}, {Geisler}, {Guhathakurta}, {Hodge}, {Karachentseva}, {Sarajedini},
  \& {Seitzer}}]{2003A&A...398..467K}
{Karachentsev}, I.~D., {Sharina}, M.~E., {Dolphin}, A.~E., {et~al.} 2003, \aap,
  398, 467

\bibitem[{{Kim} {et~al.}(2015){Kim}, {Jerjen}, {Mackey}, {Da Costa}, \&
  {Milone}}]{2015ApJ...804L..44K}
{Kim}, D., {Jerjen}, H., {Mackey}, D., {Da Costa}, G.~S., \& {Milone}, A.~P.
  2015, \apjl, 804, L44

\bibitem[{{Kniazev} {et~al.}(2004){Kniazev}, {Grebel}, {Pustilnik}, {Pramskij},
  {Kniazeva}, {Prada}, \& {Harbeck}}]{2004AJ....127..704K}
{Kniazev}, A.~Y., {Grebel}, E.~K., {Pustilnik}, S.~A., {et~al.} 2004, \aj, 127,
  704

\bibitem[{{Kormendy} {et~al.}(2010){Kormendy}, {Drory}, {Bender}, \&
  {Cornell}}]{2010ApJ...723...54K}
{Kormendy}, J., {Drory}, N., {Bender}, R., \& {Cornell}, M.~E. 2010, \apj, 723,
  54

\bibitem[{{Kroupa}(2012)}]{2012PASA...29..395K}
{Kroupa}, P. 2012, \pasa, 29, 395

\bibitem[{{Kroupa} {et~al.}(2010){Kroupa}, {Famaey}, {de Boer},
  {Dabringhausen}, {Pawlowski}, {Boily}, {Jerjen}, {Forbes}, {Hensler}, \&
  {Metz}}]{2010A&A...523A..32K}
{Kroupa}, P., {Famaey}, B., {de Boer}, K.~S., {et~al.} 2010, \aap, 523, A32

\bibitem[{{Kunth} {et~al.}(1988){Kunth}, {Maurogordato}, \&
  {Vigroux}}]{1988A&A...204...10K}
{Kunth}, D., {Maurogordato}, S., \& {Vigroux}, L. 1988, \aap, 204, 10

\bibitem[{{Libeskind} {et~al.}(2014){Libeskind}, {Knebe}, {Hoffman}, \&
  {Gottl{\"o}ber}}]{2014MNRAS.443.1274L}
{Libeskind}, N.~I., {Knebe}, A., {Hoffman}, Y., \& {Gottl{\"o}ber}, S. 2014,
  \mnras, 443, 1274

\bibitem[{{L{\'o}pez-Corredoira} \& {Kroupa}(2016)}]{2016ApJ...817...75L}
{L{\'o}pez-Corredoira}, M. \& {Kroupa}, P. 2016, \apj, 817, 75

\bibitem[{{Lupton}(2005)}]{SloanConv}
{Lupton}, R. 2005, Transformations between SDSS magnitudes and other systems
  https://www.sdss3.org/dr10/algorithms/sdssUBVRITransform.php/

\bibitem[{{Makarov} {et~al.}(2013){Makarov}, {Makarova}, \&
  {Uklein}}]{2013AstBu..68..125M}
{Makarov}, D.~I., {Makarova}, L.~N., \& {Uklein}, R.~I. 2013, Astrophysical
  Bulletin, 68, 125

\bibitem[{{McConnachie}(2012)}]{2012AJ....144....4M}
{McConnachie}, A.~W. 2012, AJ, 144, 4

\bibitem[{{McConnachie} {et~al.}(2009){McConnachie}, {Irwin}, {Ibata},
  {Dubinski}, {Widrow}, {Martin}, {C{\^o}t{\'e}}, {Dotter}, {Navarro},
  {Ferguson}, {Puzia}, {Lewis}, {Babul}, {Barmby}, {Bienaym{\'e}}, {Chapman},
  {Cockcroft}, {Collins}, {Fardal}, {Harris}, {Huxor}, {Mackey},
  {Pe{\~n}arrubia}, {Rich}, {Richer}, {Siebert}, {Tanvir}, {Valls-Gabaud}, \&
  {Venn}}]{2009Natur.461...66M}
{McConnachie}, A.~W., {Irwin}, M.~J., {Ibata}, R.~A., {et~al.} 2009, \nat, 461,
  66

\bibitem[{{McQuinn} {et~al.}(2016){McQuinn}, {Skillman}, {Dolphin}, {Berg}, \&
  {Kennicutt}}]{2016ApJ...826...21M}
{McQuinn}, K.~B.~W., {Skillman}, E.~D., {Dolphin}, A.~E., {Berg}, D., \&
  {Kennicutt}, R. 2016, \apj, 826, 21

\bibitem[{{Merritt} {et~al.}(2014){Merritt}, {van Dokkum}, \&
  {Abraham}}]{2014ApJ...787L..37M}
{Merritt}, A., {van Dokkum}, P., \& {Abraham}, R. 2014, \apjl, 787, L37

\bibitem[{{Merritt} {et~al.}(2016){Merritt}, {van Dokkum}, {Danieli},
  {Abraham}, {Zhang}, {Karachentsev}, \& {Makarova}}]{2016ApJ...833..168M}
{Merritt}, A., {van Dokkum}, P., {Danieli}, S., {et~al.} 2016, \apj, 833, 168

\bibitem[{{Mihos} {et~al.}(2013){Mihos}, {Harding}, {Spengler}, {Rudick}, \&
  {Feldmeier}}]{2013ApJ...762...82M}
{Mihos}, J.~C., {Harding}, P., {Spengler}, C.~E., {Rudick}, C.~S., \&
  {Feldmeier}, J.~J. 2013, \apj, 762, 82

\bibitem[{{M{\"u}ller} {et~al.}(2015){M{\"u}ller}, {Jerjen}, \&
  {Binggeli}}]{2015A&A...583A..79M}
{M{\"u}ller}, O., {Jerjen}, H., \& {Binggeli}, B. 2015, \aap, 583, A79

\bibitem[{{M{\"u}ller} {et~al.}(2017){M{\"u}ller}, {Jerjen}, \&
  {Binggeli}}]{2017A&A...597A...7M}
{M{\"u}ller}, O., {Jerjen}, H., \& {Binggeli}, B. 2017, \aap, 597, A7

\bibitem[{{M{\"u}ller} {et~al.}(2016){M{\"u}ller}, {Jerjen}, {Pawlowski}, \&
  {Binggeli}}]{2016A&A...595A.119M}
{M{\"u}ller}, O., {Jerjen}, H., {Pawlowski}, M.~S., \& {Binggeli}, B. 2016,
  \aap, 595, A119

\bibitem[{{Nataf}(2015)}]{2015MNRAS.449.1171N}
{Nataf}, D.~M. 2015, \mnras, 449, 1171

\bibitem[{{Ordenes-Brice{\~n}o} {et~al.}(2016){Ordenes-Brice{\~n}o}, {Taylor},
  {Puzia}, {Mu{\~n}oz}, {Eigenthaler}, {Georgiev}, {Goudfrooij}, {Hilker},
  {Lan{\c c}on}, {Mamon}, {Mieske}, {Miller}, {Peng}, \&
  {S{\'a}nchez-Janssen}}]{2016MNRAS.463.1284O}
{Ordenes-Brice{\~n}o}, Y., {Taylor}, M.~A., {Puzia}, T.~H., {et~al.} 2016,
  \mnras, 463, 1284

\bibitem[{{Papaderos} {et~al.}(1996{\natexlab{a}}){Papaderos}, {Loose},
  {Fricke}, \& {Thuan}}]{1996A&A...314...59P}
{Papaderos}, P., {Loose}, H.-H., {Fricke}, K.~J., \& {Thuan}, T.~X.
  1996{\natexlab{a}}, \aap, 314, 59

\bibitem[{{Papaderos} {et~al.}(1996{\natexlab{b}}){Papaderos}, {Loose},
  {Thuan}, \& {Fricke}}]{1996A&AS..120..207P}
{Papaderos}, P., {Loose}, H.-H., {Thuan}, T.~X., \& {Fricke}, K.~J.
  1996{\natexlab{b}}, \aaps, 120, 207

\bibitem[{{Paturel} {et~al.}(2003){Paturel}, {Petit}, {Prugniel}, {Theureau},
  {Rousseau}, {Brouty}, {Dubois}, \& {Cambr{\'e}sy}}]{2003A&A...412...45P}
{Paturel}, G., {Petit}, C., {Prugniel}, P., {et~al.} 2003, \aap, 412, 45

\bibitem[{{Pawlowski} {et~al.}(2014){Pawlowski}, {Famaey}, {Jerjen}, {Merritt},
  {Kroupa}, {Dabringhausen}, {L{\"u}ghausen}, {Forbes}, {Hensler}, {Hammer},
  {Puech}, {Fouquet}, {Flores}, \& {Yang}}]{2014MNRAS.442.2362P}
{Pawlowski}, M.~S., {Famaey}, B., {Jerjen}, H., {et~al.} 2014, \mnras, 442,
  2362

\bibitem[{{Pawlowski} {et~al.}(2015){Pawlowski}, {Famaey}, {Merritt}, \&
  {Kroupa}}]{2015ApJ...815...19P}
{Pawlowski}, M.~S., {Famaey}, B., {Merritt}, D., \& {Kroupa}, P. 2015, \apj,
  815, 19

\bibitem[{{Pawlowski} {et~al.}(2013){Pawlowski}, {Kroupa}, \&
  {Jerjen}}]{2013MNRAS.435.1928P}
{Pawlowski}, M.~S., {Kroupa}, P., \& {Jerjen}, H. 2013, \mnras, 435, 1928

\bibitem[{{Pawlowski} {et~al.}(2012){Pawlowski}, {Pflamm-Altenburg}, \&
  {Kroupa}}]{2012MNRAS.423.1109P}
{Pawlowski}, M.~S., {Pflamm-Altenburg}, J., \& {Kroupa}, P. 2012, \mnras, 423,
  1109

\bibitem[{{Rekola} {et~al.}(2005){Rekola}, {Jerjen}, \&
  {Flynn}}]{2005A&A...437..823R}
{Rekola}, R., {Jerjen}, H., \& {Flynn}, C. 2005, \aap, 437, 823

\bibitem[{{Rizzi} {et~al.}(2007){Rizzi}, {Held}, {Saviane}, {Tully}, \&
  {Gullieuszik}}]{2007MNRAS.380.1255R}
{Rizzi}, L., {Held}, E.~V., {Saviane}, I., {Tully}, R.~B., \& {Gullieuszik}, M.
  2007, \mnras, 380, 1255

\bibitem[{{Sand} {et~al.}(2014){Sand}, {Crnojevi{\'c}}, {Strader}, {Toloba},
  {Simon}, {Caldwell}, {Guhathakurta}, {McLeod}, \&
  {Seth}}]{2014ApJ...793L...7S}
{Sand}, D.~J., {Crnojevi{\'c}}, D., {Strader}, J., {et~al.} 2014, \apjl, 793,
  L7

\bibitem[{{Sersic}(1968)}]{1968adga.book.....S}
{Sersic}, J.~L. 1968, {Atlas de galaxias australes}

\bibitem[{{Smith Castelli} {et~al.}(2016){Smith Castelli}, {Faifer}, \&
  {Escudero}}]{2016A&A...596A..23S}
{Smith Castelli}, A.~V., {Faifer}, F.~R., \& {Escudero}, C.~G. 2016, \aap, 596,
  A23

\bibitem[{{Taylor} {et~al.}(2016{\natexlab{a}}){Taylor}, {Mu{\~n}oz}, {Puzia},
  {Mieske}, {Eigenthaler}, \& {Bovill}}]{2016arXiv160807285T}
{Taylor}, M.~A., {Mu{\~n}oz}, R.~P., {Puzia}, T.~H., {et~al.}
  2016{\natexlab{a}}, ArXiv e-prints [\eprint[arXiv]{1608.07285}]

\bibitem[{{Taylor} {et~al.}(2016{\natexlab{b}}){Taylor}, {Puzia}, {Mu{\~n}oz},
  {Mieske}, {Zhang}, {Eigenthaler}, \& {Bovill}}]{2016arXiv160807288T}
{Taylor}, M.~A., {Puzia}, T.~H., {Mu{\~n}oz}, R.~P., {et~al.}
  2016{\natexlab{b}}, ArXiv e-prints [\eprint[arXiv]{1608.07288}]

\bibitem[{{Tikhonov} {et~al.}(2015){Tikhonov}, {Lebedev}, \&
  {Galazutdinova}}]{2015AstL...41..239T}
{Tikhonov}, N.~A., {Lebedev}, V.~S., \& {Galazutdinova}, O.~A. 2015, Astronomy
  Letters, 41, 239

\bibitem[{{Tonry} {et~al.}(2001){Tonry}, {Dressler}, {Blakeslee}, {Ajhar},
  {Fletcher}, {Luppino}, {Metzger}, \& {Moore}}]{2001ApJ...546..681T}
{Tonry}, J.~L., {Dressler}, A., {Blakeslee}, J.~P., {et~al.} 2001, \apj, 546,
  681

\bibitem[{{Toomre} \& {Toomre}(1972)}]{1972BAAS....4..214T}
{Toomre}, A. \& {Toomre}, J. 1972, in \baas, Vol.~4, Bulletin of the American
  Astronomical Society, 214

\bibitem[{{Trentham} \& {Tully}(2002)}]{2002MNRAS.335..712T}
{Trentham}, N. \& {Tully}, R.~B. 2002, \mnras, 335, 712

\bibitem[{{Tully}(1988)}]{1988ngc..book.....T}
{Tully}, R.~B. 1988, {Nearby galaxies catalog}

\bibitem[{{Tully} {et~al.}(2015){Tully}, {Libeskind}, {Karachentsev},
  {Karachentseva}, {Rizzi}, \& {Shaya}}]{2015ApJ...802L..25T}
{Tully}, R.~B., {Libeskind}, N.~I., {Karachentsev}, I.~D., {et~al.} 2015,
  \apjl, 802, L25

\bibitem[{{van Dokkum} {et~al.}(2014){van Dokkum}, {Abraham}, \&
  {Merritt}}]{2014ApJ...782L..24V}
{van Dokkum}, P.~G., {Abraham}, R., \& {Merritt}, A. 2014, \apjl, 782, L24

\bibitem[{{Welker} {et~al.}(2014){Welker}, {Devriendt}, {Dubois}, {Pichon}, \&
  {Peirani}}]{2014MNRAS.445L..46W}
{Welker}, C., {Devriendt}, J., {Dubois}, Y., {Pichon}, C., \& {Peirani}, S.
  2014, \mnras, 445, L46

\bibitem[{{York} {et~al.}(2000){York}, {Adelman}, {Anderson}, {Anderson},
  {Annis}, {Bahcall}, {Bakken}, {Barkhouser}, {Bastian}, {Berman}, {Boroski},
  {Bracker}, {Briegel}, {Briggs}, {Brinkmann}, {Brunner}, {Burles}, {Carey},
  {Carr}, {Castander}, {Chen}, {Colestock}, {Connolly}, {Crocker}, {Csabai},
  {Czarapata}, {Davis}, {Doi}, {Dombeck}, {Eisenstein}, {Ellman}, {Elms},
  {Evans}, {Fan}, {Federwitz}, {Fiscelli}, {Friedman}, {Frieman}, {Fukugita},
  {Gillespie}, {Gunn}, {Gurbani}, {de Haas}, {Haldeman}, {Harris}, {Hayes},
  {Heckman}, {Hennessy}, {Hindsley}, {Holm}, {Holmgren}, {Huang}, {Hull},
  {Husby}, {Ichikawa}, {Ichikawa}, {Ivezi{\'c}}, {Kent}, {Kim}, {Kinney},
  {Klaene}, {Kleinman}, {Kleinman}, {Knapp}, {Korienek}, {Kron}, {Kunszt},
  {Lamb}, {Lee}, {Leger}, {Limmongkol}, {Lindenmeyer}, {Long}, {Loomis},
  {Loveday}, {Lucinio}, {Lupton}, {MacKinnon}, {Mannery}, {Mantsch}, {Margon},
  {McGehee}, {McKay}, {Meiksin}, {Merelli}, {Monet}, {Munn}, {Narayanan},
  {Nash}, {Neilsen}, {Neswold}, {Newberg}, {Nichol}, {Nicinski}, {Nonino},
  {Okada}, {Okamura}, {Ostriker}, {Owen}, {Pauls}, {Peoples}, {Peterson},
  {Petravick}, {Pier}, {Pope}, {Pordes}, {Prosapio}, {Rechenmacher}, {Quinn},
  {Richards}, {Richmond}, {Rivetta}, {Rockosi}, {Ruthmansdorfer}, {Sandford},
  {Schlegel}, {Schneider}, {Sekiguchi}, {Sergey}, {Shimasaku}, {Siegmund},
  {Smee}, {Smith}, {Snedden}, {Stone}, {Stoughton}, {Strauss}, {Stubbs},
  {SubbaRao}, {Szalay}, {Szapudi}, {Szokoly}, {Thakar}, {Tremonti}, {Tucker},
  {Uomoto}, {Vanden Berk}, {Vogeley}, {Waddell}, {Wang}, {Watanabe},
  {Weinberg}, {Yanny}, {Yasuda}, \& {SDSS Collaboration}}]{2000AJ....120.1579Y}
{York}, D.~G., {Adelman}, J., {Anderson}, Jr., J.~E., {et~al.} 2000, \aj, 120,
  1579

\end{thebibliography}

\end{document}